\documentclass[pdflatex,sn-mathphys-num]{sn-jnl}

\usepackage{graphicx}
\usepackage{subcaption}
\usepackage{import}
\usepackage{transparent}
\usepackage{float}
\usepackage{placeins}
\usepackage{url}
\usepackage{siunitx}
\usepackage{comment}
\usepackage{booktabs}
\usepackage{amsmath,amssymb,amsfonts}
\usepackage{rotating}
\usepackage{makecell}
\usepackage{array}
\usepackage{xcolor}

\let\orcidlogo\relax
\usepackage{orcidlink}

\newcommand{\nucleus}{\textsc{Nucleus}}
\newcommand{\cresst}{\textsc{Cresst}}
\newcommand{\tesseract}{\textsc{Tesseract}}
\newcommand{\stella}{\textsc{Stella}}
\newcommand{\cawo}{CaWO$_4$}

\newcommand{\cevns}{CE$\nu$NS}
\DeclareSIUnit{\piece}{piece}
\makeatletter

\newcommand{\breakafterfirstauthor}{%
    \g@addto@macro\artauthors{%
        \par
        \let\authorsep\@empty
        \advance\punctcount by -1\relax
    }%
}

\newcommand{\nucleuscollaboration}{%
  \g@addto@macro\artauthors{%
    \\[0.3em]\textnormal{(the \nucleus\ collaboration)}%
  }%
}

\makeatother

\begin{document}

\title[Development and Commissioning of the Cryogenic Target Detector for \nucleus]{Development and Commissioning of the Cryogenic Target Detectors for the Technical Run of the \nucleus\ Experiment}

\renewcommand{\Authorfont}{%
    \fontsize{9bp}{11bp}\selectfont
    \boldmath
    \titraggedcenter
}

\renewcommand{\addressfont}{%
    \fontsize{7bp}{9bp}\selectfont
    \boldmath
    \titraggedcenter
}




%
\newcount\authorStyle

\authorStyle=1

%
\newcount\instByName

\instByName=0

\newcommand{\iNUCLEUScontactEmail}{Contact E-Mail of NUCLEUS Collaboration : info@nucleus-experiment.org}

\ifcase\authorStyle
    \collaboration{NUCLEUS Collaboration}
\else
\fi

\ifnum\authorStyle=1
\else
    \newcommand{\orgdiv}[1]{#1}%
    \newcommand{\orgname}[1]{#1}%
    \newcommand{\orgaddress}[1]{#1}%
    \newcommand{\street}[1]{#1}%
    \newcommand{\postcode}[1]{#1}%
    \newcommand{\city}[1]{#1}%
    \newcommand{\state}[1]{#1}%
    \newcommand{\country}[1]{#1}%
\fi





\newcommand{\iMBI}{%
    \orgname{Marietta-Blau-Institut f{\"u}r Teilchenphysik der {\"O}sterreichischen Akademie der Wissenschaften}, 
    \orgaddress{
        \street{Dominikanerbastei~16}, 
        \city{Wien}, 
        \postcode{A-1010}, 
        \country{Austria}%
        }%
    }

\newcommand{\iTUW}{%
    \orgdiv{Atominstitut}, 
    \orgname{Technische Universit\"at Wien}, 
    \orgaddress{
        \street{Stadionallee~2}, 
        \city{Wien}, 
        \postcode{A-1020}, 
        \country{Austria}%
        }%
    }


\newcommand{\iCEA}{%
    \orgdiv{IRFU}, 
    \orgname{CEA, Universit\'{e} Paris-Saclay}, 
    \orgaddress{
        \street{B\^{a}timent 141}, 
        \city{Gif-sur-Yvette}, 
        \postcode{F-91191}, 
        \country{France}%
        }%
    }

\newcommand{\iEdF}{%
    \orgdiv{Centre nucl{\'e}aire de production d'{\'}electricit{\'e} de Chooz, Service Automatismes-Essais}, 
    \orgname{{\'E}lectricit{\'e} de France}, 
    \orgaddress{
        \street{}, 
        \city{Givet}, 
        \postcode{F-08600}, 
        \country{France}%
        }%
    }


\newcommand{\iMPIK}{%
    \orgname{Max-Planck-Institut für Kernphysik}, 
    \orgaddress{%
        \street{Saupfercheckweg 1}, 
        \city{Heidelberg}, 
        \postcode{D-69117}, 
        \country{Germany}%
        }%
    }

\newcommand{\iMPP}{%
    \orgname{Max-Planck-Institut f{\"u}r Physik}, 
    \orgaddress{
        \street{Boltzmannstra{\ss}e~8}, 
        \city{Garching}, 
        \postcode{D-85748}, 
        \country{Germany}%
        }%
    }

\newcommand{\iTUM}{%
    \orgdiv{Physik-Department, TUM School of Natural Sciences}, 
    \orgname{Technische Universit\"at M\"unchen}, 
    \orgaddress{
        \street{James-Franck-Straße 1}, 
        \city{Garching}, 
        \postcode{D-85748}, 
        \country{Germany}%
        }%
    }


\newcommand{\iINFNRoma}{%
    \orgname{Istituto Nazionale di Fisica Nucleare -- Sezione di Roma}, 
    \orgaddress{
        \street{Piazzale Aldo Moro 2}, 
        \city{Roma}, 
        \postcode{I-00185}, 
        \country{Italy}%
        }%
    }

\newcommand{\iSapienza}{%
    \orgdiv{Dipartimento di Fisica}, 
    \orgname{Sapienza Universit\`{a} di Roma}, 
    \orgaddress{
        \street{Piazzale Aldo Moro 5}, 
        \city{Roma}, 
        \postcode{I-00185}, 
        \country{Italy}%
        }%
    }
    
\newcommand{\iINFNTorVergata}{%
    \orgname{Istituto Nazionale di Fisica Nucleare -- Sezione di Roma "Tor Vergata"}, 
    \orgaddress{
        \street{Via della Ricerca Scientifica 1}, 
        \city{Roma}, 
        \postcode{I-00133}, 
        \country{Italy}%
        }%
    }

\newcommand{\iTorVergata}{%
    \orgdiv{Dipartimento di Fisica}, 
    \orgname{Universit\`{a} di Roma "Tor Vergata"}, 
    \orgaddress{
        \street{Via della Ricerca Scientifica 1}, 
        \city{Roma}, 
        \postcode{I-00133}, 
        \country{Italy}%
        }%
    }

\newcommand{\iCNR}{%
    \orgdiv{Istituto di Nanotecnologia}, 
    \orgname{Consiglio Nazionale delle Ricerche}, 
    \orgaddress{
        \street{Piazzale Aldo Moro 5}, 
        \city{Roma}, 
        \postcode{I-00185}, 
        \country{Italy}%
        }%
    }

\newcommand{\iINFNFerrara}{%
    \orgname{Istituto Nazionale di Fisica Nucleare -- Sezione di Ferrara}, 
    \orgaddress{
        \street{Via Giuseppe Saragat 1c}, 
        \city{Ferrara}, 
        \postcode{I-44122}, 
        \country{Italy}%
        }%
    }

\newcommand{\iFerrara}{%
    \orgdiv{Dipartimento di Fisica}, 
    \orgname{Universit{\`a} di Ferrara}, 
    \orgaddress{
        \street{Via Giuseppe Saragat 1}, 
        \city{Ferrara}, 
        \postcode{I-44122}, 
        \country{Italy}%
        }%
    }

\newcommand{\iINFNLnGS}{%
    \orgname{Istituto Nazionale di Fisica Nucleare -- Laboratori Nazionali del Gran Sasso}, 
    \orgaddress{
        \street{Via Giovanni Acitelli 22}, 
        \city{Assergi (L’Aquila)}, 
        \postcode{I-67100}, 
        \country{Italy}%
        }%
    }

\newcommand{\iBicocca}{%
    \orgdiv{Dipartimento di Fisica}, 
    \orgname{Universit\`{a} di Milano Bicocca}, 
    \orgaddress{
        \street{}, 
        \city{Milan}, 
        \postcode{I-20126}, 
        \country{Italy}%
        }%
    }


\newcommand{\iCoimbra}{%
    \orgdiv{LIBPhys-UC, Departamento de Fisica}, 
    \orgname{Universidade de Coimbra}, 
    \orgaddress{
        \street{Rua Larga 3004-516}, 
        \city{Coimbra}, 
        \postcode{P3004-516}, 
        \country{Portugal}%
        }%
    }

%
%
\newcommand{\icorrespond}{Corresponding author: nicole.schermer@tum.de }
\newcommand{\iAlsoAtCoimbra}{Also at \iCoimbra}
\newcommand{\iNowAtMPIK}{Now at \iMPIK}
\newcommand{\iNowAtLNGS}{Now at \iINFNLnGS}
\newcommand{\iNowAtKiutra}{Now at kiutra GmbH, Fl{\"o}{\ss}ergasse 2, D-81369 Munich, Germany}


\ifnum\instByName=1
    
    \newcommand{\MBI}{MBI}
    \newcommand{\TUW}{TUW}

    \newcommand{\CEA}{CEA}
    \newcommand{\EdF}{EdF}

    \newcommand{\MPIK}{MPIK}
    \newcommand{\MPP}{MPP}
    \newcommand{\TUM}{TUM}

    \newcommand{\INFNRoma}{INFNRoma}
    \newcommand{\Sapienza}{Sapienza}
    \newcommand{\INFNTorVergata}{INFNTorVergata}
    \newcommand{\TorVergata}{TorVergata}
    \newcommand{\CNR}{CNR}
    \newcommand{\INFNFerrara}{INFNFerrara}
    \newcommand{\Ferrara}{Ferrara}
    \newcommand{\INFNLnGS}{INFNLnGS}
    \newcommand{\Bicocca}{Bicocca}    

    \newcommand{\Coimbra}{Coimbra}
\else    
    
    \newcommand{\MBI}{3}
    \newcommand{\TUW}{1}

    \newcommand{\CEA}{5}
    \newcommand{\EdF}{ERROR}

    \newcommand{\MPIK}{ERROR}
    \newcommand{\MPP}{2}
    \newcommand{\TUM}{8}

    \newcommand{\INFNRoma}{6}
    \newcommand{\Sapienza}{7}
    
    \newcommand{\INFNTorVergata}{4}
    \newcommand{\TorVergata}{9}
    
    \newcommand{\CNR}{ERROR}
    
    \newcommand{\Ferrara}{10}
    \newcommand{\INFNFerrara}{11}
    
    \newcommand{\INFNLnGS}{ERROR}
    
    \newcommand{\Bicocca}{ERROR}    

    \newcommand{\Coimbra}{ERROR}

\fi




\ifcase\authorStyle
    \author{N.~Schermer~\orcidlink{0009-0004-4213-5154}}
    \affiliation{\iTUM}
\or
    \author[\TUM]{
        \fnm{N.} 
        \sur{Schermer} 
        \orcidlink{0009-0004-4213-5154}
        \textsuperscript{\correspondingauthor,\,}
        }

\else
\fi

\breakafterfirstauthor

\ifcase\authorStyle
    \author{H.~Abele~\orcidlink{0000-0002-6832-9051}}
    \affiliation{\iTUW}
\or
    \author[\TUW]{
        \fnm{H.} 
        \sur{Abele} 
        \orcidlink{0000-0002-6832-9051}
        }
\else
\fi

\ifcase\authorStyle
    \author{G.~Angloher}
     \affiliation{\iMPP}
\or
    \author[\MPP]{
    \fnm{G.} 
    \sur{Angloher} 
    }
\else
\fi

\ifcase\authorStyle
    \author{B.~Arnold}
    \affiliation{\iMBI}
\or
    \author[\MBI]{
        \fnm{B.}
        \sur{Arnold}
        }
\else
\fi
        
\ifcase\authorStyle
    \author{M.~Atzori~Corona~\orcidlink{0000-0001-5092-3602}}
    \affiliation{\iINFNTorVergata}
\or
    \author[\INFNTorVergata]{
    \fnm{M.}
    \sur{Atzori~Corona}
    \orcidlink{0000-0001-5092-3602}
    }
\else
\fi

\ifcase\authorStyle
    \author{A.~Bento~\orcidlink{0000-0002-3817-6015}}
    \thanks{\iAlsoAtCoimbra}
     \affiliation{\iMPP}
\or
    \author[\MPP]{
        \fnm{A.}
        \sur{Bento}
        \orcidlink{0000-0002-3817-6015}
        \textsuperscript{\symAlsoAtCoimbra,\,}%
        }
    \else
\fi

\ifcase\authorStyle
    \author{E.~Bossio~\orcidlink{0000-0001-9304-1829}}
    \affiliation{\iCEA}
\or
    \author[\CEA]{
        \fnm{E.}
        \sur{Bossio}
        \orcidlink{0000-0001-9304-1829}
        }
    \else
\fi
    
\ifcase\authorStyle
    \author{F.~Buchsteiner}
    \affiliation{\iMBI}
\or
    \author[\MBI]{
        \fnm{F.}
        \sur{Buchsteiner}
        }
    \else
\fi
    
\ifcase\authorStyle
    \author{J.~Burkhart~\orcidlink{0000-0002-1989-7845}}
    \affiliation{\iMBI}
\or
    \author[\MBI]{
        \fnm{J.}
        \sur{Burkhart}
        \orcidlink{0000-0002-1989-7845}
        }
\else
\fi
    

\ifcase\authorStyle
    \author{F.~Cappella~\orcidlink{0000-0003-0900-6794}}
    \affiliation{\iINFNRoma}
\or
    \author[\INFNRoma]{
        \fnm{F.}
        \sur{Cappella}
        \orcidlink{0000-0003-0900-6794}
        }
\else
\fi
    
\ifcase\authorStyle
    \author{M.~Cappelli~\orcidlink{0009-0002-6148-5964}}
    \affiliation{\iSapienza}
    \affiliation{\iINFNRoma}
\or
    \author[\Sapienza, \INFNRoma]{
        \fnm{M.}
        \sur{Cappelli}
        \orcidlink{0009-0002-6148-5964}
        }
\else
\fi
    

\ifcase\authorStyle
    \author{N.~Casali~\orcidlink{0000-0003-3669-8247}}
    \affiliation{\iINFNRoma}
\or
    \author[\INFNRoma]{
        \fnm{N.}
        \sur{Casali}
        \orcidlink{0000-0003-3669-8247}
        }
\else
\fi
    
\ifcase\authorStyle
    \author{R.~Cerulli~\orcidlink{0000-0003-2051-3471}}
    \affiliation{\iINFNTorVergata}
\or
    \author[\INFNTorVergata]{
        \fnm{R.}
        \sur{Cerulli}
        \orcidlink{0000-0003-2051-3471}
        }
\else
\fi
    

\ifcase\authorStyle
    \author{A.~Cruciani~\orcidlink{0000-0003-2247-8067}}
    \affiliation{\iINFNRoma}
\or
    \author[\INFNRoma]{
        \fnm{A.}
        \sur{Cruciani}
        \orcidlink{0000-0003-2247-8067}
        }
\else
\fi
    
\ifcase\authorStyle
    \author{G.~Del~Castello~\orcidlink{0000-0001-7182-358X}}
    \affiliation{\iINFNRoma}
\or
    \author[\INFNRoma]{
        \fnm{G.}
        \sur{Del~Castello}
        \orcidlink{0000-0001-7182-358X}
        }
\else
\fi


\ifcase\authorStyle
    \author{S.~Dorer~\orcidlink{0009-0001-1670-5780}}
    \affiliation{\iTUW}
\or
    \author[\TUW]{
        \fnm{S.}
        \sur{Dorer}
        \orcidlink{0009-0001-1670-5780}
        }
\else
\fi
    
\ifcase\authorStyle
    \author{A.~Erhart~\orcidlink{0000-0002-8721-177X}}
    \affiliation{\iTUM}
\or
    \author[\TUM]{
        \fnm{A.}
        \sur{Erhart}
        \orcidlink{0000-0002-8721-177X}
        }
\else
\fi
    
\ifcase\authorStyle
    \author{M.~Friedl~\orcidlink{0000-0002-7420-2559}}
    \affiliation{\iMBI}
\or
    \author[\MBI]{
        \fnm{M.}
        \sur{Friedl}
        \orcidlink{0000-0002-7420-2559}}
\else
\fi

\ifcase\authorStyle
    \author{S.~Fichtinger}
    \affiliation{\iMBI}
\or
    \author[\MBI]{
        \fnm{S.}
        \sur{Fichtinger}
        }
\else
\fi


\ifcase\authorStyle
    \author{V.M.~Ghete~\orcidlink{0000-0002-9595-6560}}
    \affiliation{\iMBI}
\or
    \author[\MBI]{
        \fnm{V.M.}
        \sur{Ghete}
        \orcidlink{0000-0002-9595-6560}
        }
\else
\fi

\ifcase\authorStyle
    \author{M.~Giammei~\orcidlink{0009-0006-9104-2055}}
    \affiliation{\iTorVergata}
    \affiliation{\iINFNTorVergata}
\or
    \author[\TorVergata, \INFNTorVergata]{
        \fnm{M.}
        \sur{Giammei}
        \orcidlink{0009-0006-9104-2055}
        }
\else
\fi


\ifcase\authorStyle
    \author{J.~Hakenm{\"u}ller~\orcidlink{0000-0003-0470-3320}}
    \affiliation{\iMBI}
\or
    \author[\MBI]{
        \fnm{J.}
        \sur{Hakenm{\"u}ller}
        \orcidlink{0000-0003-0470-3320}}
\else
\fi

\ifcase\authorStyle
    \author{D.~Hauff}
     \affiliation{\iMPP}
    \affiliation{\iTUM}
\or
    \author[\MPP, \TUM]{
        \fnm{D.}
        \sur{Hauff}
        }
\else
\fi

\ifcase\authorStyle
    \author{M.~Hock}
    \affiliation{\iTUM}
\or
    \author[\TUM]{
        \fnm{M.}
        \sur{Hock}
        }
\else
\fi
    
\ifcase\authorStyle
    \author{F.~Jeanneau~\orcidlink{0000-0002-6360-6136}}
    \affiliation{\iCEA}
\or
    \author[\CEA]{
        \fnm{F.}
        \sur{Jeanneau}
        \orcidlink{0000-0002-6360-6136}}
\else
\fi

\ifcase\authorStyle
    \author{E.~Jericha~\orcidlink{0000-0002-8663-0526}}
    \affiliation{\iTUW}
\or
    \author[\TUW]{
        \fnm{E.}
        \sur{Jericha}
        }
\else
\fi

\ifcase\authorStyle
    \author{M.~Kaznacheeva~\orcidlink{0000-0002-2712-1326}}
    \affiliation{\iTUM}
\or
    \author[\TUM]{
        \fnm{M.}
        \sur{Kaznacheeva}
        \orcidlink{0000-0002-2712-1326}
        }
\else
\fi


\ifcase\authorStyle
    \author{H.~Kluck~\orcidlink{0000-0003-3061-3732}}
    \affiliation{\iMBI}
\or
\author[\MBI]{
    \fnm{H.} 
    \sur{Kluck} 
    \orcidlink{0000-0003-3061-3732}
    }
\else
\fi

\ifcase\authorStyle
    \author{A.~Langenk{\"a}mper}
     \affiliation{\iMPP}
\or
    \author[\MPP]{
        \fnm{A.} 
        \sur{Langenk\"{a}mper} 
        \orcidlink{}
        }
\else
\fi

\ifcase\authorStyle
    \author{T.~Lasserre~\orcidlink{0000-0002-4975-2321}}
    \thanks{\iNowAtMPIK}
    \affiliation{\iCEA}
    \affiliation{\iTUM}
\or
    \author[\CEA, \TUM]{
        \fnm{T.} 
        \sur{Lasserre} 
        \orcidlink{0000-0002-4975-2321}
        \textsuperscript{\symNowAtMPIK,\,}%
        }
\else
\fi

\ifcase\authorStyle
    \author{D.~Lhuillier~\orcidlink{0000-0003-2324-0149}}
    \affiliation{\iCEA}
\or
    \author[\CEA]{
        \fnm{D.} 
        \sur{Lhuillier} 
        \orcidlink{0000-0003-2324-0149}
        }
\else
\fi

\ifcase\authorStyle
    \author{M.~Mancuso~\orcidlink{0000-0001-9805-475X}}
     \affiliation{\iMPP}
\or
    \author[\MPP]{
        \fnm{M.} 
        \sur{Mancuso} 
        \orcidlink{0000-0001-9805-475X}
        }
\else
\fi

\ifcase\authorStyle
    \author{B.~Mauri}
     \affiliation{\iMPP}
\or
    \author[\MPP]{
        \fnm{B.} 
        \sur{Mauri} 
        }
\else
\fi

\ifcase\authorStyle
    \author{A.~Mazzolari}
    \affiliation{\iFerrara}
    \affiliation{\iINFNFerrara}
\or
    \author[\Ferrara, \INFNFerrara]{
        \fnm{A.} 
        \sur{Mazzolari} 
        }
\else
\fi

\ifcase\authorStyle
    \author{L.~McCallin}
    \affiliation{\iCEA}
\or
    \author[\CEA]{
        \fnm{L.} 
        \sur{McCallin} 
        }
\else
\fi


\ifcase\authorStyle
    \author{H.~Neyrial}
    \affiliation{\iCEA}
\or
    \author[\CEA]{
        \fnm{H.} 
        \sur{Neyrial} 
        }
\else
\fi

\ifcase\authorStyle
    \author{C.~Nones}
    \affiliation{\iCEA}
\or
    \author[\CEA]{
        \fnm{C.} 
        \sur{Nones} 
        }
\else
\fi

\ifcase\authorStyle
    \author{L.~Oberauer}
    \affiliation{\iTUM}
\or
    \author[\TUM]{
        \fnm{L.} 
        \sur{Oberauer} 
        }
\else
\fi

\ifcase\authorStyle
    \author{L.~Peters~\orcidlink{0000-0002-1649-8582}}
    \thanks{\iNowAtMPIK}
    \affiliation{\iTUM}
    \affiliation{\iCEA}
\or
    \author[\TUM, \CEA]{
        \fnm{L.} 
        \sur{Peters} 
        \orcidlink{0000-0002-1649-8582}
        \textsuperscript{\symNowAtMPIK,\,}%
        }
\else
\fi

\ifcase\authorStyle
    \author{F.~Petricca~\orcidlink{0000-0002-6355-2545}}
    \affiliation{\iMPP}
\or
    \author[\MPP]{
        \fnm{F.} 
        \sur{Petricca} 
        \orcidlink{0000-0002-6355-2545}
        }
\else
\fi

\ifcase\authorStyle
    \author{W.~Potzel}
    \affiliation{\iTUM}
\or
    \author[\TUM]{
        \fnm{W.} 
        \sur{Potzel} 
        }
\else
\fi

\ifcase\authorStyle
    \author{F.~Pr\"{o}bst}
     \affiliation{\iMPP}
\or
    \author[\MPP]{
        \fnm{F.} 
        \sur{Pr\"{o}bst} 
        }
\else
\fi

\ifcase\authorStyle
    \author{F.~Reindl~\orcidlink{0000-0003-0151-2174}}
    \affiliation{\iMBI}
    \affiliation{\iTUW}
\or
    \author[\MBI, \TUW]{
        \fnm{F.} 
        \sur{Reindl} 
        \orcidlink{0000-0003-0151-2174}
        }
\else
\fi


\ifcase\authorStyle
    \author{M.~Romagnoni}
    \affiliation{\iFerrara}
    \affiliation{\iINFNFerrara}
\or
    \author[\Ferrara, \INFNFerrara]{
        \fnm{M.} 
        \sur{Romagnoni} 
        }
\else
\fi

\ifcase\authorStyle
    \author{J.~Rothe~\orcidlink{0000-0001-5748-7428}}
    \thanks{\iNowAtKiutra}
    \affiliation{\iTUM}
\or
    \author[\TUM]{
        \fnm{J.} 
        \sur{Rothe} 
        \orcidlink{0000-0001-5748-7428}
        \textsuperscript{\symNowAtKiutra,\,}%
        }
\else
\fi

\ifcase\authorStyle
    \author{J.~Schieck~\orcidlink{0000-0002-1058-8093}}
    \affiliation{\iMBI}
    \affiliation{\iTUW}
\or
    \author[\MBI, \TUW]{
        \fnm{J.} 
        \sur{Schieck} 
        \orcidlink{0000-0002-1058-8093}
        }
\else
\fi

\ifcase\authorStyle
    \author{S.~Sch\"{o}nert~\orcidlink{0000-0001-5276-2881}}
    \affiliation{\iTUM}
\or
    \author[\TUM]{
        \fnm{S.} 
        \sur{Sch\"{o}nert} 
        \orcidlink{0000-0001-5276-2881}
        }
\else
\fi

\ifcase\authorStyle
    \author{C.~Schwertner}
    \affiliation{\iMBI}
    \affiliation{\iTUW}
\or
    \author[\MBI, \TUW]{
        \fnm{C.} 
        \sur{Schwertner} 
        }
\else
\fi

\ifcase\authorStyle
    \author{L.~Scola}
    \affiliation{\iCEA}
\or
    \author[\CEA]{
        \fnm{L.} 
        \sur{Scola} 
        }
\else
\fi

\ifcase\authorStyle
    \author{L.~Stodolsky}
     \affiliation{\iMPP}
\or
    \author[\MPP]{
        \fnm{L.} 
        \sur{Stodolsky} 
        }
\else
\fi

\ifcase\authorStyle
    \author{A.~Schr{\"o}der~\orcidlink{0009-0005-1598-1635}}
    \affiliation{\iTUM}
\or
    \author[\TUM]{
        \fnm{A.} 
        \sur{Schr{\"o}der} 
        \orcidlink{0009-0005-1598-1635}
        }
\else
\fi

\ifcase\authorStyle
    \author{R.~Strauss~\orcidlink{0000-0002-5589-9952}}
    \affiliation{\iTUM}
\or
    \author[\TUM]{
        \fnm{R.} 
        \sur{Strauss} 
        \orcidlink{0000-0002-5589-9952}
        }
\else
\fi


\ifcase\authorStyle
    \author{R.~Thalmeier~\orcidlink{0009-0003-4480-0990}}
    \affiliation{\iMBI}
\or
    \author[\MBI]{
        \fnm{R.} 
        \sur{Thalmeier} 
        \orcidlink{0009-0003-4480-0990}
        }
\else
\fi

\ifcase\authorStyle
    \author{C.~Tomei}
    \affiliation{\iINFNRoma}
\or
    \author[\INFNRoma]{
        \fnm{C.} 
        \sur{Tomei} 
        }
\else
\fi

\ifcase\authorStyle
    \author{L.~Valla~\orcidlink{0009-0003-7140-9196}}
    \affiliation{\iMBI}
\or
    \author[\MBI]{
        \fnm{L.} 
        \sur{Valla} 
        \orcidlink{0009-0003-7140-9196}
        }
\else
\fi

\ifcase\authorStyle
    \author{M.~Vignati~\orcidlink{0000-0002-8945-1128}}
    \affiliation{\iSapienza}
    \affiliation{\iINFNRoma}
\or
    \author[\Sapienza, \INFNRoma]{
        \fnm{M.} 
        \sur{Vignati} 
        \orcidlink{0000-0002-8945-1128}
        }
\else
\fi

\ifcase\authorStyle
    \author{M.~Vivier~\orcidlink{0000-0003-2199-0958}}
    \affiliation{\iCEA}
\or
    \author[\CEA]{
        \fnm{M.} 
        \sur{Vivier} 
        \orcidlink{0000-0003-2199-0958}
        }
\else
\fi


\ifcase\authorStyle
    \author{A.~Wallach~\orcidlink{ 0009-0009-1703-9634}}
    \affiliation{\iTUM}
\or
    \author[\TUM]{
        \fnm{A.} 
        \sur{Wallach} 
        \orcidlink{0009-0009-1703-9634}
        }
\else
\fi

\ifcase\authorStyle
    \author{P.~Wasser~\orcidlink{0009-0004-7650-7307}}
    \affiliation{\iTUM}
\or
    \author[\TUM]{
        \fnm{P.} 
        \sur{Wasser} 
        \orcidlink{0009-0004-7650-7307}
        }
\else
\fi

\ifcase\authorStyle
    \author{L.~Wienke~\orcidlink{0009-0006-5548-2109}}
    \affiliation{\iTUM}
\or
    \author[\TUM]{
        \fnm{L.} 
        \sur{Wienke} 
        \orcidlink{0009-0006-5548-2109}
        }
\else
\fi

\nucleuscollaboration

\ifcase\authorStyle
\or


    \affil[\TUW]{\iTUW}

    \affil[\MPP]{\iMPP}
    
    
    \affil[\MBI]{\iMBI}

    \affil[\INFNTorVergata]{\iINFNTorVergata}

    \affil[\CEA]{\iCEA}

    \affil[\INFNRoma]{\iINFNRoma}

    \affil[\Sapienza]{\iSapienza}

    \affil[\TUM]{\iTUM}

    \affil[\TorVergata]{\iTorVergata}

    \affil[\Ferrara]{\iFerrara}

    \affil[\INFNFerrara]{\iINFNFerrara}

    

    %

    \affil[ ]{\vspace{1ex}} 

    \newcommand{\correspondingauthor}{*}
    \affil[\correspondingauthor]{\icorrespond}

    \newcommand{\symAlsoAtCoimbra}{\dag}
    \affil[\symAlsoAtCoimbra]{\iAlsoAtCoimbra}
    
    
    \newcommand{\symNowAtMPIK}{\ddag}
    \affil[\symNowAtMPIK]{\iNowAtMPIK}



    \newcommand{\symNowAtKiutra}{\textdollar}
    \affil[\symNowAtKiutra]{\iNowAtKiutra}

    
    
    
\else
\fi

\abstract{
    The \nucleus\ experiment aims to study coherent elastic neutrino--nucleus scattering (\cevns) of reactor electron antineutrinos at the Chooz nuclear power plant in France. In this work, the cryogenic target-detector module for the \nucleus\ Technical Run was developed and commissioned at the Technical University of Munich. The module comprises four gram-scale \cawo\ detectors, each equipped with two Transition Edge Sensors (TES), providing a total target mass of \SI{6.96}{g}. We present the design, integration, detector characterization, and X-ray-based energy calibration of these detectors. Across six characterized detectors, a mean baseline resolution of $\overline{\sigma}_{\mathrm{BL}}=\SI{2.79 \pm 0.60}{eV}$ was achieved. The best-performing detector reached $\sigma_{\mathrm{BL}}=(2.16\pm0.02_{\mathrm{stat}})$~\si{\electronvolt}, surpassing the design goal by a factor of two and representing a state-of-the-art result for a cryogenic \cawo\ detector. The target detector module was successfully operated simultaneously with the surrounding Cryogenic Outer Veto, which consists of six kg-scale Ge detectors for background discrimination, showing no measurable cross-talk. These results demonstrate the readiness of the target-detector system for the \nucleus\ Technical Run at Chooz and mark a key milestone in the development of cryogenic detectors for reactor-\cevns\ measurements.}


\maketitle
\section{Introduction}
\label{sec:introduction}

Coherent elastic neutrino--nucleus scattering (\cevns) is a neutral-current interaction, in which a neutrino scatters coherently off an entire nucleus at low momentum transfer~\cite{Freedman:1973yd}. Although this Standard Model process has a comparatively large cross section, its experimental detection is challenging since the resulting nuclear recoils have very low energies. COHERENT first observed \cevns\ in 2017~\cite{COHERENT:2017ipa} and subsequently confirmed the process with measurements using different target materials~\cite{COHERENT:2020iec, COHERENT:2024axu}. Recent reactor-based results further demonstrate the potential to measure \cevns\ from reactor electron antineutrinos~\cite{Ackermann:2025obx}. Here, however, the nuclear recoil energies lie predominantly well below \SI{1}{keV}, making detectors with thresholds of only a few tens of electronvolts essential to fully exploit the potential of a precision \cevns\ measurement~\cite{2026prospectnucleusexperimentchooz-xzr1}.

The \nucleus\ experiment addresses this challenge with gram-scale cryogenic detectors based on Calcium Tungstate (\cawo) absorber crystals equipped with superconducting Transition Edge Sensors (TES) achieving ultra-low energy thresholds of \SI{20}{eV}. To obtain a measurable reactor-\cevns\ event rate, several gram-scale detectors are combined into a compact cryogenic detector module.~\cite{nucleus2017}

The first deployment at a reactor site of the cryogenic target detectors will take place during the \textit{Technical Run}~\cite{2026prospectnucleusexperimentchooz-xzr1} of the \nucleus\ experiment at the Chooz nuclear power plant in France. This run represents an intermediate step toward the first reactor-\cevns\ physics measurement. Its main goals are to demonstrate the stable operation of the cryogenic target detectors, together with the cryogenic infrastructure, shielding, and veto systems, under reactor-site conditions; to confirm that the achieved detector energy resolution can be maintained in situ at the reactor site; and to characterize the background environment at Chooz.

The cryogenic target detector module developed for the Technical Run is referred to in this work as the \textit{Minimal Detector Module} (MDM). The designation reflects its deliberately compact design, which builds on well-established detector-holder concepts demonstrated in Refs.~\cite{LBR_paper,Abele:2026sqm, xrf-paper} and avoids the more complex holding mechanisms foreseen for the Physics Run~\cite{nucleus2019, Schermer_Nicole_2025_detectors_proceeding}. It therefore represents the simplest configuration of multiple target detectors compatible with the limited space inside the surrounding veto detectors. The MDM accommodates four $5 \times 5 \times 11.5 $~\si{\milli\meter^3} \cawo\ detectors with double-TES readout, providing a total target mass of approximately \SI{7}{g}.

An important challenge for this first reactor measurement is the \textit{Low-Energy Excess} (LEE), a yet unexplained event population at very low energies observed in several cryogenic experiments~\cite{Baxter_2025,Fuss:2022fxe}. 
During the Technical Run, the LEE is expected to dominate the sub-keV spectrum of the target detectors, making the characterization and understanding of low-energy background contributions a central objective.
Nevertheless, even in the presence of the LEE, by exploiting both energy and reactor-power time dependence, \nucleus\ is expected to achieve competitive sensitivity to several beyond-the-Standard-Model scenarios already during the Technical Run, including light vector mediators and the neutrino magnetic moment, even in the absence of a Standard Model \cevns\ observation~\cite{2026prospectnucleusexperimentchooz-xzr1}. Strategies for improved LEE discrimination are foreseen in later experimental phases~\cite{2026prospectnucleusexperimentchooz-xzr1}.

\begin{figure}[htbp]
    \centering
    \includegraphics[width=\textwidth]{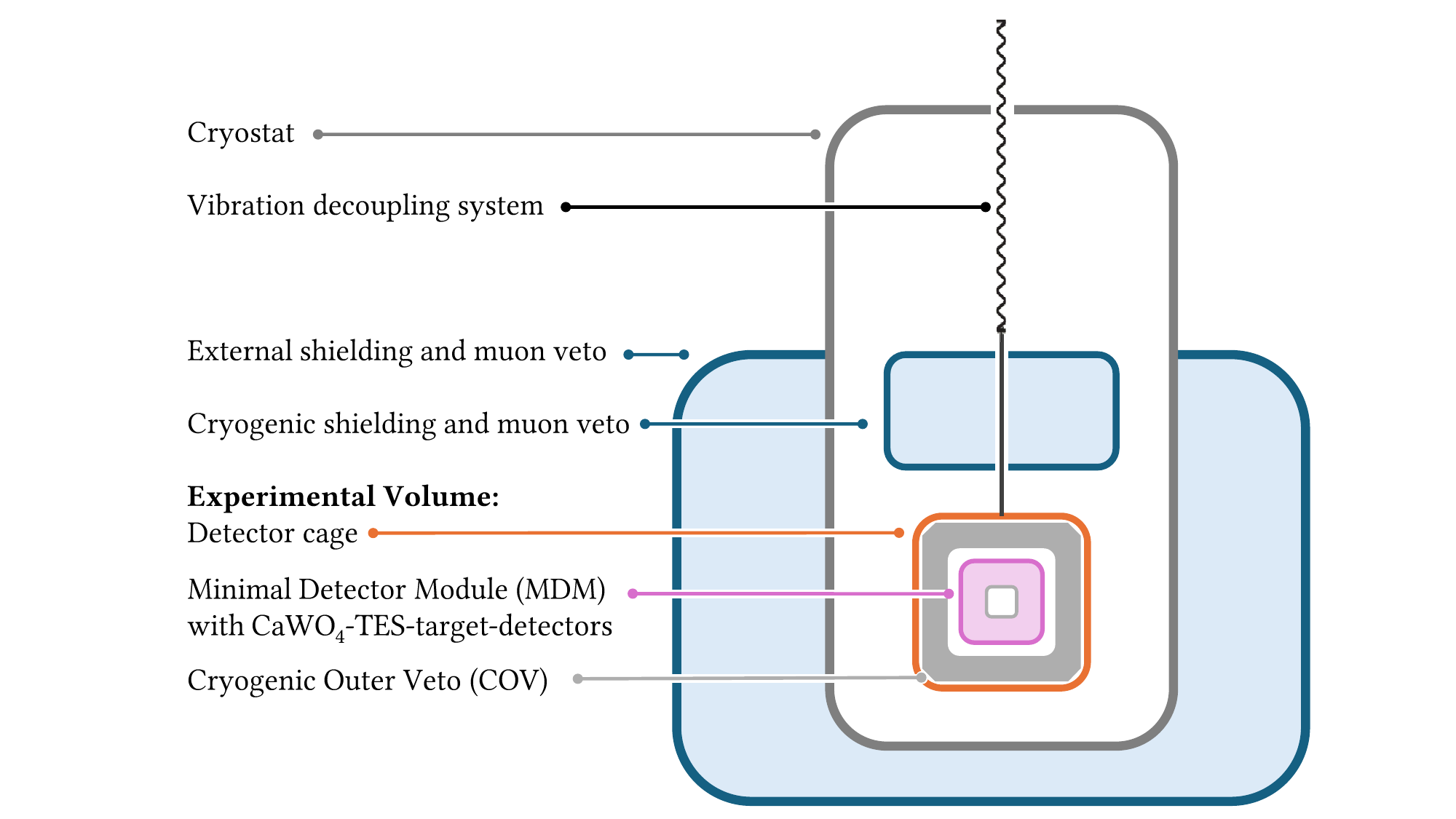}
    \caption[Schematic drawing of NUCLEUS setup]
    {
    Schematic overview of the \nucleus\ experimental setup, including the cryostat, vibration-decoupling system, shielding and veto systems, as well as the experimental volume consisting of the detector cage, the Cryogenic Outer Veto, and the Minimal Detector Module (MDM) with its \cawo\ TES-target detectors.
    }
    \label{fig:nucleus_setup_shematic_drawing}
\end{figure}

Fig.~\ref{fig:nucleus_setup_shematic_drawing} presents a schematic overview of the \nucleus\ experimental setup. The experimental volume is cooled to millikelvin temperatures inside the cryostat and mechanically isolated from cryostat-induced and environmental vibrations by a vibration-decoupling system~\cite{Wex:2025jwu}. Passive shielding and active muon-veto systems are installed both inside and outside the cryostat to suppress and identify external background events. Further details on the shielding and background-reduction concept are provided in Refs.~\cite{LBR_paper,nucleus_background_paper}. The experimental volume comprises the detector cage, which houses the six kg-scale Ge detectors of the COV surrounding the MDM. 

This work presents the development, construction, and commissioning of the MDM for its first deployment at the Chooz nuclear power plant. Sec.~\ref{sec:concept_setup} introduces the cryogenic detector concept, including the double-TES design, and detector production, while the details about the experimental setup, and about the cryogenic infrastructure and the SQUID-based readout, are presented in Sec.~\ref{sec:experimental_setup}. Sec.~\ref{sec:MDM_design} presents the mechanical, thermal, electrical, and calibration design of the MDM and assesses its expected contribution to the particle background during the Technical Run. The commissioning procedure and detector performance obtained at the Technical University of Munich (TUM), including the pulse-shape analysis, energy calibration, baseline-resolution results, and compatibility with the COV, are discussed in Sec.~\ref{sec:commissioning}. Finally, Sec.~\ref{sec:conclusion} summarizes the results and outlines the next steps toward a reactor-\cevns\ measurement with \nucleus.


\FloatBarrier
\section{Cryogenic Target Detector Concept}
\label{sec:concept_setup}

The cryogenic target detectors developed for the \nucleus\ reactor measurement with the MDM are gram-scale \cawo\ detectors equipped with tungsten TESs. A particle interaction in the absorber crystal deposits energy, which is ultimately converted into phonon excitations. The detector response contains a prompt athermal component and a slower thermal component arising from phonon scattering, reflections, and down-conversion in the absorber~\cite{Probst:1995hjq}. A fraction of the phonon population is absorbed in the TES, where it thermalizes and produces a small increase in the sensor temperature. The TES is operated within the transition between the superconducting and normal-conducting states, where small temperature changes lead to measurable changes in resistance.

\begin{figure}[htbp]
    \centering
    \includegraphics[width=\textwidth]{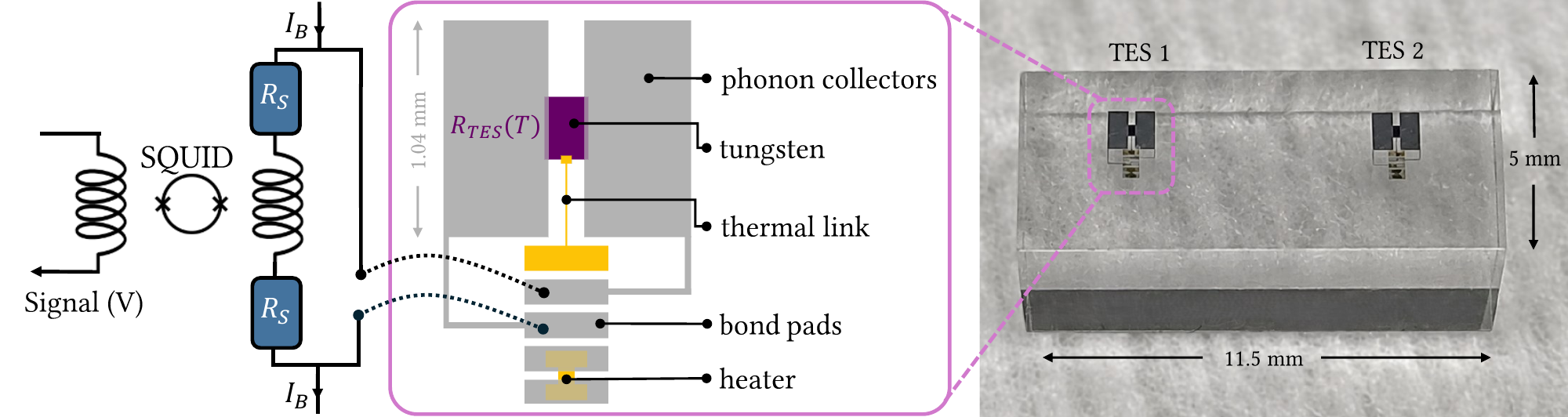}
    \caption[Double-TES detector with TES and SQUID-readout schematic]
    {
    Photograph of a \cawo\ target detector crystal with dimensions of $5\times5\times11.5~\mathrm{mm}^3$, equipped with two TESs. The schematic shows the TES layout and the SQUID-based readout principle. The TES consists of a superconducting tungsten film, aluminium phonon collectors, a gold thermal link, and an ohmic heater. Dedicated bond pads connect the TES structures to the readout circuit via wire bonds. The temperature-dependent TES resistance $R_\mathrm{TES}(T)$ is converted into a voltage signal by the SQUID-based readout system which is detailed in Sec.~\ref{sec:experimental_setup}.  
    For the double-TES readout, both TESs are read out by independent SQUID channels, while one of the two heaters heats the absorber and keeps both TESs close to their transition temperatures of about \SI{15}{mK}.
    }
    \label{fig:double_TES_shematic}
\end{figure}

Fig.~\ref{fig:double_TES_shematic} shows a photograph of one \cawo\ double-TES detector together with schematics of the TES layout, which are detailed in the following subsections.

\FloatBarrier
\subsection{\texorpdfstring{\cawo}{CaWO4} detector}
\label{subsec:gram_scale_calorimeters}

Each target detector consists of a \cawo\ absorber crystal operated as a cryogenic detector at temperatures of around \SI{15}{mK}. 
The use of \cawo\ provides heavy target nuclei, in particular tungsten, which significantly boosts the \cevns\ signal strength due to the approximately neutron-number-squared dependence of the cross section. 
In addition to the material choice, the absorber mass is a central design parameter. Gram-scale detectors are advantageous for achieving low energy thresholds required for reactor-\cevns\ detection at the low-energy frontier. 
The initial design goal of \nucleus\ was to reach energy thresholds as low as \SI{20}{eV}, enabling the detection of a sufficient number of events for a reactor-\cevns\ measurement with a detector concept employing $5\times5\times5~\si{mm^3}$ absorber cubes equipped with a single TES~\cite{strauss2017}. With the introduction of the double-TES readout, the absorber geometry was extended to provide sufficient surface area for two independent TESs while preserving the small absorber mass required for a low energy threshold.
Another advantage of using gram-scale detectors is to guarantee a sufficiently low background event rate per detector, which is particularly important for operation in surface environments such as the reactor site at Chooz. 
In the MDM, each detector has dimensions of $11.5\times5\times5~\mathrm{mm}^3$, corresponding to a mass of  \SI{1.74}{g}. For the Technical Run of \nucleus, four \cawo\ target detectors are deployed, providing a total target mass of \SI{6.96}{g}.

\subsection{TES Design}
\label{subsec:TES_design}
The TES design of the target detectors, shown schematically in Fig.~\ref{fig:double_TES_shematic}, is based on the design developed for the dark-matter search \cresst\ experiment~\cite{cresst2024} and was adapted for \nucleus\ as studied in Ref.~\cite{rothePhd}.
The TES size represents a compromise between maximizing the phonon-collection efficiency and minimizing the total heat capacity at a given heat capacity of the film.

Tungsten is used as the thermometric material because its superconducting transition temperature can be tuned to the millikelvin range required for low-threshold detector operation. While bulk tungsten has a transition temperature of about \SI{15}{mK}~\cite{Lassner1999Tungsten}, the transition temperature of thin-film tungsten TESs can be adjusted through the choice of deposition and fabrication parameters.
The tungsten film has a thickness of \SI{200}{\nano\meter} and an active area of $0.30\times0.21~\mathrm{mm}^2$. 
To increase the effective phonon collection area without significantly increasing the TES heat capacity, the tungsten film is coupled to superconducting aluminium phonon collectors~\cite{Angloher2016}. These collectors are operated far below their superconducting transition temperature of about \SI{1.2}{K}~\cite{RevModPhys.26.277}, partially overlap with the tungsten film, and mediate the transfer of athermal phonons from the absorber crystal into the TES, as can be seen from Fig.~\ref{fig:double_TES_shematic}. The aluminium film has a thickness of \SI{1}{\micro\meter} and covers an area of $0.51\times1.04~\mathrm{mm}^2$.
The TES is weakly coupled to the heat bath by a gold thermal link with a thickness of \SI{100}{\nano\meter} and dimensions of $0.01\times0.39~\mathrm{mm}^2$. 
This link defines the thermal relaxation path of the TES system and is designed to provide a sufficiently long relaxation time for efficient collection of athermal phonons, as described by the pulse-shape model in Ref.~\cite{Probst:1995hjq}, which is discussed in Sec.~\ref{sec:pulse_template}.
In addition, since the tungsten transition temperature is above the cryostat base temperature of below \SI{8}{mK}, the TES is heated into its operating point by an ohmic gold heater with a thickness of \SI{100}{\nano\meter}. 
Separate aluminium bond pads are used for the electrical connection of the phonon collectors and the ohmic heater, while a gold bond pad connects the thermal link to the heat bath. 
Typically, constant bias currents between \SI{1}{\micro\ampere} and \SI{6}{\micro\ampere} are applied during operation.

\subsection{Double-TES Detector Concept}
\label{subsec:double_TES_concept}

As shown in Fig.~\ref{fig:double_TES_shematic}, \nucleus\ uses a double-TES concept, in which two spatially separated TESs are fabricated on the same \cawo\ absorber and read out simultaneously and independently. This concept has been investigated in cryogenic dark-matter searches by \cresst~\cite{cresst2024} and \tesseract~\cite{TESSERACT:2025tfw}, and was adapted for the \nucleus\ target detectors to identify TES-related contributions to the LEE~\cite{Abele:2026sqm}. One possible contribution to the LEE is stress relaxation or other localized processes at the interfaces between the metallic sensor films and the crystal surface.
An event originating in the close vicinity of one TES is expected to produce a localized signal with a larger amplitude in the nearby sensor. In contrast, particle interactions in the absorber volume are expected to produce phonon signals shared equally between both TESs. The comparison of the two simultaneously measured TES channels, therefore, provides spatial information and can be used to distinguish bulk-like events from TES-local event populations.
Each TES of a double-TES detector is equipped with a nearby heater structure for redundancy as illustrated in Fig.~\ref{fig:double_TES_shematic}. Since the heater increases the temperature of the absorber crystal and thereby also heats both TES films, either heater can be used to establish the operating points of both TESs. This requires sufficiently similar transition temperatures of the two tungsten films, such that both TESs can be operated simultaneously within their superconducting transitions. The commissioning measurements presented in this work showed that comparable operating conditions can be achieved for both TESs, independent of which heater is used.

\subsection{TES Production and Detector Selection}
\label{subsec:TES_production_and_characterization}

\begin{figure}[htbp]
    \centering
    \begin{subfigure}[b]{0.38\columnwidth}
        \centering
        \includegraphics[width=\textwidth]{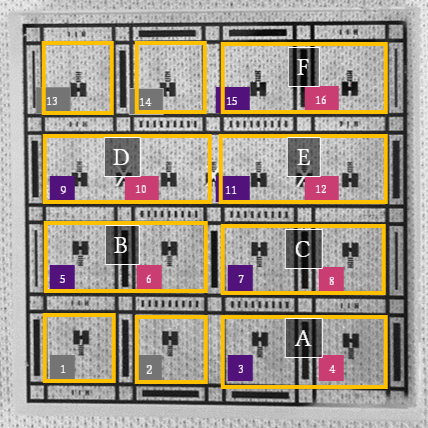}
        \caption{}
        \label{fig:transition_array}
    \end{subfigure}
    \hfill
    \begin{subfigure}[b]{0.6\columnwidth}
        \centering
        \includegraphics[width=\textwidth]{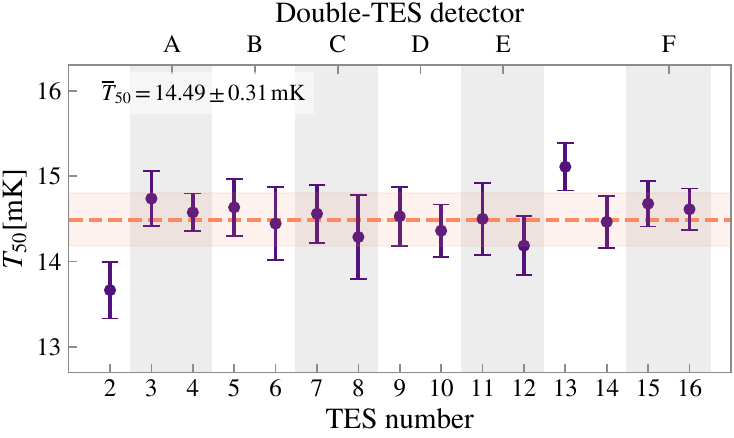}
        \caption{}
        \label{fig:transition_temperatures}
    \end{subfigure}
    \caption[CaWO$_4$ array and transition temperatures]{
    \textbf{(a)} \cawo\ crystal with 16 TES structures before dicing. The selected cutting scheme is highlighted. Single-TES detectors, double-TES pairs, and test structures are indicated separately. The crossed-out test-TES structures for detectors D and E, as well as other remaining metallic films, were chemically etched after cutting.
    \textbf{(b)} Transition temperatures $T_{50}$ of the 15 TESs, while TES~1 did not show a measurable transition. 
    The error bars represent the transition width.
    The dashed line indicates the mean transition temperature of \SI{14.49}{\milli\kelvin}, and the shaded band shows the corresponding standard deviation of \SI{0.31}{\milli\kelvin}. The selected TES pairs forming the six double-TES detectors A--F are highlighted.
    }
    \label{fig:array_transition_temperatures}
\end{figure}

The fabrication process of the TESs consists of successive deposition and lithographic patterning of tungsten, aluminium, and gold films, followed by dicing of the substrate into individual detectors and polishing. Since several TES structures are fabricated on a common absorber substrate in a single fabrication cycle, the process enables the production of multiple detectors with comparable sensor properties.
 This is based on the technology originally developed for the dark-matter-search experiment \cresst~\cite{Bravin1999CRESST,Angloher2026CRESST}.
For the detector production discussed here, 16 TES structures were deposited on a single \cawo\ crystal with dimensions of $30\times30\times5~\mathrm{mm}^3$, as shown in Fig.~\ref{fig:transition_array}. 
The transition curves were measured before dicing and the associated transition temperatures $T_\mathrm{50}$ are shown in Fig.~\ref{fig:transition_temperatures}. 

\begin{figure}[htbp]
    \centering
    \begin{minipage}[c]{0.58\textwidth}
        \centering
        \includegraphics[width=\textwidth]{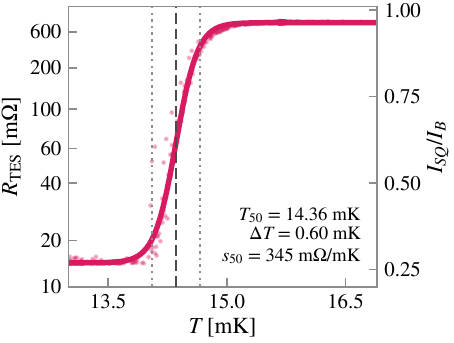}
    \end{minipage}
    \hfill
    \begin{minipage}[c]{0.4\textwidth}
        \caption[Example TES transition curve]{
        Example superconducting transition curve of TES~10, recorded with a \SI{100}{nA} bias current, and fitted with a sigmoidal function according to Equation~\ref{equ:sigmoid_fit}. 
        The dashed line represents the midpoint $T_{50}$, which defines the transition temperature.
        The dotted lines indicate the temperatures $T_{10}$ and $T_{90}$, at which the response reaches 10\% and 90\%, respectively. The difference $\Delta T = T_{90}-T_{10}$ quantifies the transition width. The slope $s_{\mathbf{50}}$ at $T_{50}$ is used as a measure of the transition steepness.
        }
        \label{fig:TES-transition_curve_example}
    \end{minipage}
\end{figure}

An example transition curve for TES~10 is shown in Fig.~\ref{fig:TES-transition_curve_example}. The transition curve is obtained from the ratio of the current through the SQUID branch, $I_{\mathrm{SQ}}$, to the applied bias current, $I_{\mathrm{B}}$. Based on the readout circuit introduced in Fig.~\ref{fig:double_TES_shematic}, this ratio can be converted into the TES resistance. 
The transition curve was recorded with a bias current of \SI{100}{\nano\ampere}, about one order of magnitude below the typical bias currents used during detector operation. This reduced current minimizes Joule heating in the TES, allowing the transition curve to be measured without significantly distorting the transition.
For characterizing the transition, the measured response is fitted with a sigmoidal function of
\begin{equation}
\label{equ:sigmoid_fit}
    f(T) = R_{\mathrm{sc}} + \frac{R_{\mathrm{nc}}-R_{\mathrm{sc}}} {1+\exp\left[-\frac{(T-T_{50})}{\delta} \right]},
\end{equation}
where $R_{\mathrm{sc}}$ and $R_{\mathrm{nc}}$ describe the superconducting and normal-conducting plateaus of the transition (see Fig.~\ref{fig:TES-transition_curve_example}), $T_{50}$ denotes the temperature at which \SI{50}{\percent} of the transition amplitude is reached, and $\delta$ is the scale parameter determining the steepness of the sigmoid. The transition width is defined as $\Delta T = T_{90}-T_{10}$, where $T_{10}$ and $T_{90}$ are the temperatures at which \SI{10}{\percent} and \SI{90}{\percent} of the transition amplitude are reached, respectively. In addition, the slope of the normalized response at the transition midpoint $s_{50} = \left.\frac{\mathrm{d}f}{\mathrm{d}T}\right|_{T=T_{50}}$ can be extracted.

For the example in Fig.~\ref{fig:TES-transition_curve_example}, the fit yields a transition temperature of $T_{50}=\SI{14.36}{mK}$, a transition width of $\Delta T =\SI{0.60}{mK}$, and a slope of $s_{50}=\SI{345}{\milli\ohm\per\milli\kelvin}$.
In general, all TES films selected for the target detectors exhibit narrow transitions with widths below \SI{1}{mK} and steep slopes comparable to those typically achieved with TESs employed in \cresst\ detectors~\cite{Abdelhameed2020Tungsten}.
A narrow, steep superconducting transition is essential for achieving high detector sensitivity, since the measured signal arises from a small temperature-induced change in TES resistance. 

In total, 15 of the 16 TESs showed a measurable superconducting transition, with an average transition temperature and the associated standard deviation of $T_{50} = (14.49 \pm 0.31)\,\si{\milli\kelvin}$. This demonstrates the homogeneity of the fabrication process and enables the selection of matched TES pairs. For the double-TES detectors, TESs with closely matched transition temperatures were paired. This is advantageous because both TESs on one absorber can then be operated simultaneously within their superconducting transition using a common heater. Based on the transition-temperature measurements, and their spatial arrangement, twelve TESs were selected and paired to form six double-TES detectors, labeled A--F in Fig.~\ref{fig:array_transition_temperatures}. The remaining TES structures, including the one without a measurable transition, were retained as single-TES detectors or test structures. After dicing, the detectors were ground and polished to their final dimensions, and residual metallic films on the crystal surfaces were removed by chemical etching. For the Technical Run at the reactor site, the MDM deploys four of the six characterized detectors.

\FloatBarrier
\section{Experimental Setup and Cryogenic Infrastructure}
\label{sec:experimental_setup}

\begin{figure}[htbp]
    \centering
    \includegraphics[width=\textwidth]{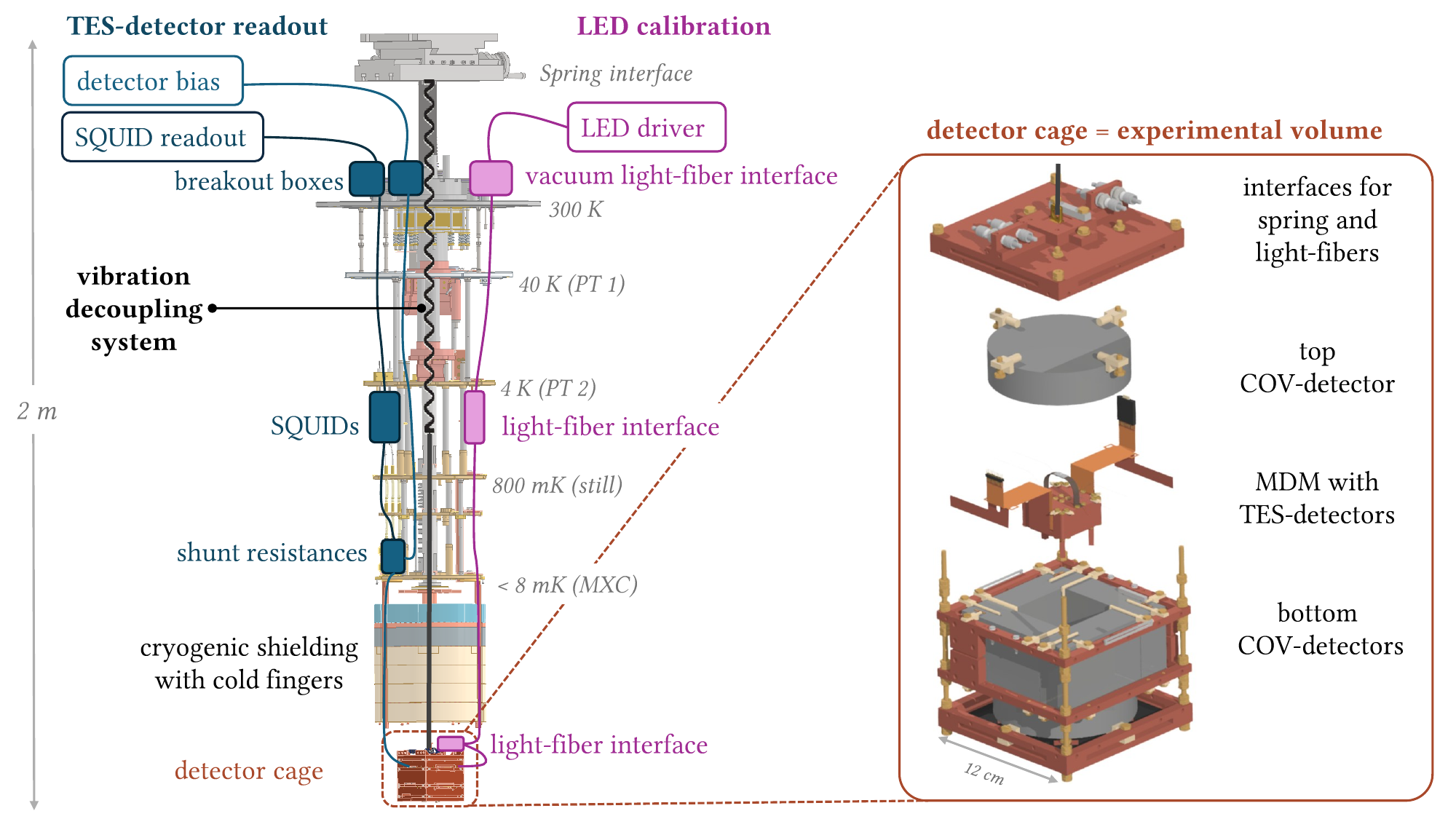}
    \caption[Schematic overview of the NUCLEUS cryogenic detector setup]{
    Schematic overview of the \nucleus\ cryogenic detector setup for the Technical Run. The target detectors of the MDM are operated inside the detector cage at millikelvin temperatures and are surrounded by the Cryogenic Outer Veto (COV). The detector cage is mechanically decoupled from pulse-tube and environmental vibrations by the vibration decoupling system and thermally connected to the mixing chamber (MXC) via cold fingers and flexible copper braids. The schematic also indicates the SQUID-based TES readout and the optical-fiber system used for LED calibration of the target detectors.
    }
    \label{fig:nucleus_setup_schematic}
\end{figure}

The target detectors are operated inside the cryogenic infrastructure of the \nucleus\ experiment developed for the Technical Run. This infrastructure provides the millikelvin environment, mechanical decoupling, SQUID-based TES readout, and in-situ calibration structures required for stable operation of the target detectors. A schematic overview of the setup is shown in Fig.~\ref{fig:nucleus_setup_schematic}.
Details about the cryogenic infrastructures can be found in Refs.~\cite{ErhartPhD,wexPhd,goupyPhd,SchermerThesisPrep2026}.

The experimental volume is located inside the detector cage, which provides a cubic volume with a side length of approximately \SI{12}{cm}. It hosts the MDM, which contains the cryogenic TES-based target detectors that require a temperature of around \SI{10}{mK}. The MDM is surrounded by the COV, which consists of six high-purity germanium crystals: two cylindrical crystals with a diameter of \SI{100}{mm} and a thickness of \SI{25}{mm}, and four rectangular crystals with dimensions of ${75 \times 25 \times 50}\si{mm^3}$. Aluminium electrodes are evaporated onto the top and bottom surfaces of each crystal, allowing the COV detectors to be read out through ionization channels.
 It provides nearly $4\pi$ coverage around the target volume and is used to identify and reject gamma- and neutron-induced background events~\cite{goupyPhd,wexPhd}.

\paragraph{Cryostat, vibration decoupling system and shielding}

Cryogenic operation is provided by a BlueFors LD400 dilution refrigerator, reaching base temperatures below \SI{8}{mK} with a cooling power of \SI{17.5}{\micro\watt} at \SI{20}{mK}~\cite{blueforsCryostat}. The system is pre-cooled to \SI{4}{K} by a Cryomech PT-415 pulse tube operating at \SI{1.4}{Hz}. To reduce the impact of pulse-tube-induced and environmental vibrations on detector performance, the detector cage is mechanically decoupled by a \SI{1.8}{m}-long spring pendulum mounted on an independent reference frame~\cite{Wex:2025jwu}. The top of the detector cage provides the mechanical interface to this vibration decoupling system, as indicated in Fig.~\ref{fig:nucleus_setup_schematic}.

The detector cage is thermally anchored to the mixing chamber of the cryostat via copper cold fingers and flexible copper braids, enabling operation of the target detectors and the COV below \SI{10}{mK}. The cold fingers bridge the distance between the mixing chamber and the detector cage, which is required by the cryogenic shielding geometry. A full description of the shielding and veto systems inside and outside the cryostat is given in Refs.~\cite{nucleus2019,LBR_paper}. These systems combine passive shielding components for neutron and gamma suppression with active veto systems for muon identification.

\paragraph{SQUID-based readout system}

As indicated in Fig.~\ref{fig:double_TES_shematic}, the signals of the target detectors are measured using a SQUID-based current readout. The temperature-dependent TES resistance  $R_\mathrm{TES}(T)$ is operated in parallel with a shunt branch, which consists of two shunt resistances of $R_\mathrm{S}=\SI{20}{\milli\ohm}$ in series with the input coil of the SQUID. A constant bias current $I_\mathrm{B}$ is applied to the circuit and splits between the TES and the shunt branch according to their resistances.
A change in TES resistance modifies the current flowing through the shunt branch. This current variation is inductively coupled to the SQUID loop, where the resulting flux change is converted into a measurable voltage response~\cite{rothePhd}. Each TES of a double-TES detector is connected to an independent readout channel, allowing the two sensor signals to be recorded simultaneously.

The DC-SQUIDs used in this setup are of the type \textit{SQ100 LTS sensors} supplied by STAR Cryoelectronics~\cite{StarCryo_SQ100} and are coupled to the \SI{4}{K} stage of the cryostat as shown in Fig.~\ref{fig:nucleus_setup_schematic}. They are connected to the shunt resistors at the mixing-chamber stage, which, in turn, are connected to the TES-detectors in the detector cage. The resulting signals, as well as the detector bias lines, are fed out via dedicated breakout boxes on top of the cryostat. All TES signals are digitized by the Versatile Data Acquisition (VDAQ) system~\cite{vdaq3,LBR_paper}, which also provides the bias and heater-control signals required for TES operation and stabilization.

\paragraph{Optical calibration system}

To provide optical calibration and stability monitoring of the target detectors, an LED calibration system is installed inside the cryostat as indicated in Fig.~\ref{fig:nucleus_setup_schematic}. The LEDs are driven by a room-temperature LED driver~\cite{DelCastello:2024ehl,DelCastello:2025tyu}, and the light is guided to the target detectors through optical fibers. The fiber system is segmented for installation and thermalization at different cryostat stages, including a vacuum feedthrough, an interface at the \SI{4}{K} stage, routing along the cold fingers, and a final interface at the detector cage. The fibers have diameters down to \SI{120}{\micro\meter} and span a total length of more than \SI{2}{m} between room temperature and the detector cage below \SI{10}{mK}. One optical fiber is deployed for each target detector, enabling independent in-situ LED calibration during dedicated calibration periods.

\FloatBarrier
\section{Detector Module Design and Background Assessment}
\label{sec:MDM_design}

This section describes the design and integration of the cryogenic target detector module developed for the \nucleus\ Technical Run, as well as the expected background contribution of its components. The MDM accommodates four double-TES \cawo\ target detectors and provides the mechanical, thermal, electrical, and calibration interfaces required for their operation within the COV in the detector cage. The design builds on a copper detector holder developed during previous R\&D and commissioning measurements at TUM~\cite{Schermer_Nicole_2025_detectors_proceeding,LBR_paper}. The module has to fit into the limited experimental volume inside the COV, use radiopure materials close to the detectors, provide reliable TES signal routing, and maintain thermal contact with the detector cage while avoiding direct contact with the surrounding veto crystals.

\subsection{Detector Holder Concept}
\begin{figure}[htbp]
    \centering
    \includegraphics[width=\textwidth]{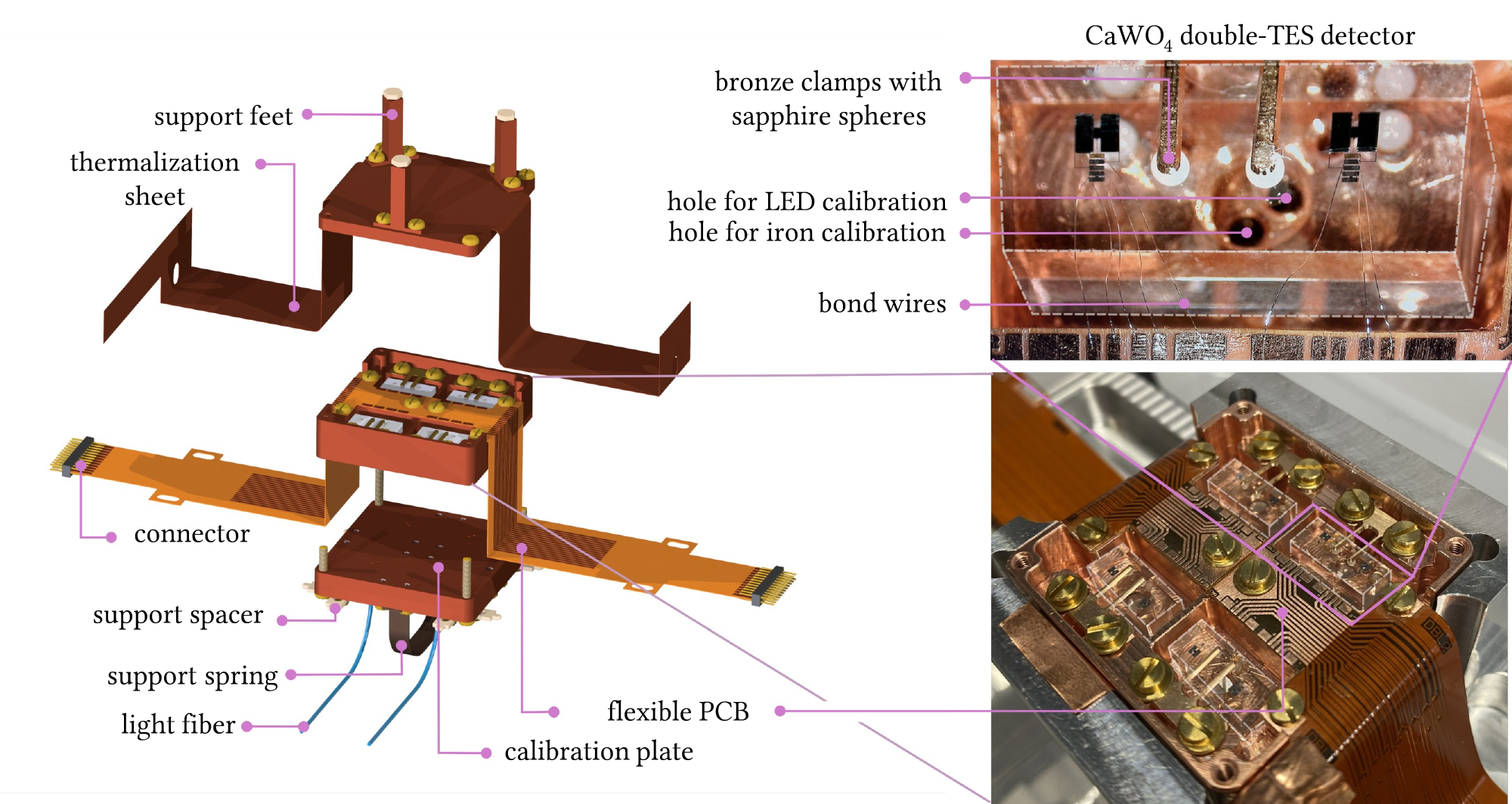}
    \caption[Exploded CAD drawing and photographs of the detector module]{
    Exploded CAD view and photographs of the MDM for the Technical Run. 
    The CAD view on the left shows the copper holder, flexible copper--Kapton PCBs, copper thermalization sheet, interchangeable calibration plates, and mechanical interfaces to the surrounding COV crystals. The photographs on the right show the four installed \cawo\ double-TES detectors and a close-up of one detector fixed by bronze clamps with \SI{1}{mm} sapphire-sphere contacts. Holes below the detectors in the calibration plate provide access for collimated $^{55}$Fe sources or optical fibers for calibration. The TESs are wire-bonded to flexible PCBs which connect to the cryostat's SQUID-based readout system. After detector mounting and wire bonding, the holder is rotated for installation of the calibration interfaces and subsequent mounting inside the COV.
    }
    \label{fig:MDM_detector_module_zoom}
\end{figure}

Fig.~\ref{fig:MDM_detector_module_zoom} gives an overview of the MDM components, which are detailed in the following. 

\paragraph{Mechanical holder and COV integration.}
The MDM housing is fabricated from NOSV copper, a high-conductivity copper alloy produced by Aurubis~\cite{aurubis} and commonly used in cryogenic applications. It accommodates four double-TES detectors in a compact geometry compatible with the surrounding COV. 
The available space inside the COV defines a maximum volume of approximately $5\times5\times5~\mathrm{cm}^3$ for the target detector module. To avoid direct contact with the COV crystals, the MDM footprint is reduced to approximately $4\times4~\mathrm{cm}^2$ and supported by defined Teflon contact points.
Teflon supports electrically isolate the MDM from the surrounding COV and prevent copper components from contacting the COV electrodes, thereby avoiding interference between the two readout systems.
As can be seen in Fig.~\ref{fig:MDM_detector_module_zoom}, these include support feet with Teflon attachments, Teflon support spacers and a bronze support spring covered with Teflon.
The copper housing largely encloses the detectors to suppress environmental infrared radiation.
Each detector is secured by bronze clamps as shown in Fig.~\ref{fig:MDM_detector_module_zoom}. 
Electrical and thermal isolation from the copper holder and the clamp is provided by sapphire spheres of \SI{1}{mm} diameter placed between the detector, the clamps, and the holder base. These contact points mechanically fix the detectors while limiting direct phonon leakage into the holder.

\paragraph{Electrical routing}
Electrical connection is realized with flexible copper--Kapton PCBs, produced by Multi-CB~\cite{MultiCB_Flex}, which guide the TES and heater signals through the \SI{2}{mm} gaps between the COV crystals to soldered connectors interfacing with the cryostat's SQUID-readout. 
Conventional FR4-core PCBs were avoided because radioactive contamination had been observed in previous studies~\cite{goupyPhd,Schermer_Nicole_2025_detectors_proceeding}. 
The PCB sections attached to the module are glued onto \SI{0.5}{mm}-thick copper plates and mechanically fixed to the holder base, providing a stable wire-bonding surface. Aluminium bond wires with a diameter of \SI{17}{\micro\meter} provide the electrical contacts to the TES and heater structures, whereas gold bond wires with a diameter of \SI{25}{\micro\meter} provide the thermal connection to the heat bath.

\paragraph{Thermalization}
The thermal conductance of the flexible PCBs is limited by the \SI{35}{\micro\meter} copper thickness and was found to be insufficient as the primary thermal link for the detector module. Therefore, a dedicated, high-purity, oxygen-free copper sheet, supplied by Goodfellow~\cite{Goodfellow_CV00_FL_000130}, is attached to the bottom of the module and extends toward the detector cage, forming the primary thermal link to the cryostat. The thermal conductance of this copper sheet was measured to be $k = \SI{38.4 \pm 3.9}{\micro\watt\per\kelvin}$ at \SI{10}{mK}, exceeding that of the PCB traces by more than an order of magnitude. Its geometry follows the PCB routing to preserve the compact module design as illustrated in Fig.~\ref{fig:MDM_detector_module_zoom}.
The copper--Kapton PCBs and thermalization components are additionally insulated to prevent contact with the COV electrodes and avoid mutual interference between the two detector systems.

\paragraph{Calibration interfaces}
Dedicated calibration interfaces are integrated into the MDM. 
As shown in Fig.~\ref{fig:MDM_detector_module_zoom}, two calibration plates below each detector, each containing holes with a diameter of \SI{1}{mm}, allow for the deployment of different calibration strategies. A light-fiber plate enables the positioning of \SI{120}{\micro\meter}-diameter optical fibers above the detectors. 
For the measurements presented in this paper, each detector was additionally equipped with an individually collimated $^{55}$Fe calibration source, yielding a rate of approximately \SI{0.03}{cps}, which results in a negligible dead time in the target detectors. The Technical Run configuration at the Chooz nuclear power plant will rely solely on LED-based calibration, as described in Sec.~\ref{sec:experimental_setup}.

\subsection{Expected Background Contributions in the Technical Run}
\label{sec:background_simulation}

The contribution of the MDM to the overall background expected during the Technical Run was evaluated using Geant4-based background simulations~\cite{geant4_2003,geant4_2016} following the procedure described in Ref.~\cite{nucleus_background_paper}. The simulations include radioactive contaminations of the detector module and surrounding components, as well as external background sources relevant to the experimental site at the nuclear reactor. Their purpose is to determine whether radioactive contaminations in materials located close to the target detectors constitute a dominant contribution to the expected low-energy background.

Table~\ref{tab:mdm_material_screening} in the Appendix summarizes the material-screening results for the MDM components and their immediate surroundings. The measurements were performed at the \stella\ (SubTErranean Low-Level Assay) facility at the Laboratori Nazionali del Gran Sasso of INFN~\cite{Laubenstein:2017yjj}. The measured activities and upper limits were used as input to the background simulations.

\begin{figure}[htbp]
    \centering
    \begin{minipage}[c]{0.55\textwidth}
        \centering
        \includegraphics[width=\textwidth]{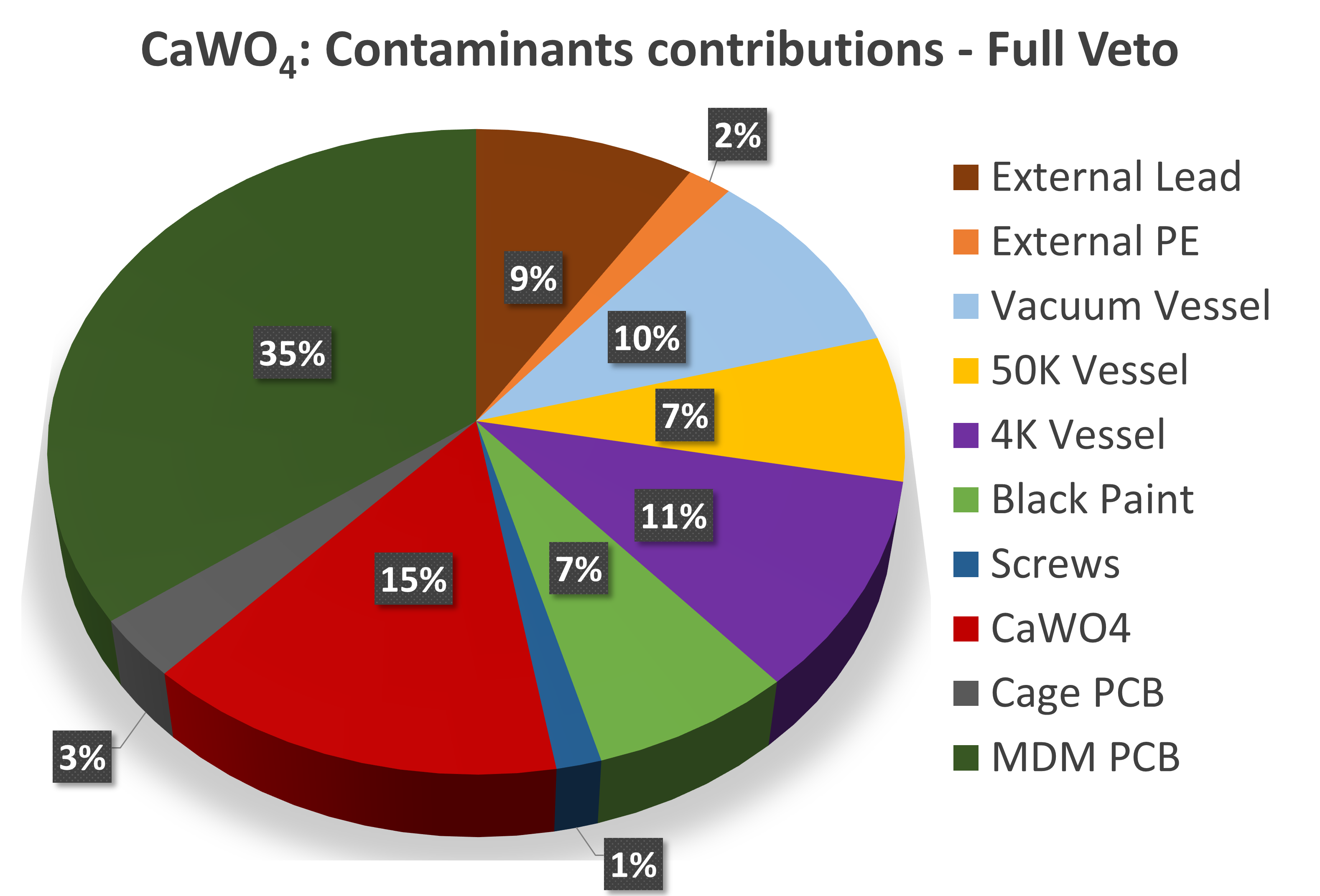}
    \end{minipage}
    \hfill
    \begin{minipage}[c]{0.42\textwidth}
        \caption[Radioactive background contributions in the Technical Run]{
        Simulated relative contributions to the background induced by radioactive contamination of the different components in the MDM and in the surrounding cryogenic environment in the \cawo\ target detectors during the \nucleus\ Technical Run. The contributions are integrated over the expected accessible energy range of \([0,10]\,\si{\kilo\electronvolt}\) 
        after applying the COV anticoincidence cut.
        }
        \label{fig:simulation_pie_chart}
    \end{minipage}
\end{figure}

Fig.~\ref{fig:simulation_pie_chart} shows the simulated relative contributions of each setup component to the background caused by radioactive contamination in the \cawo\ target detectors during the Technical Run. The contributions are integrated over the expected accessible energy range of \([0,10]\,\si{\kilo\electronvolt}\) and are shown after applying the COV anticoincidence cuts. 
Among the radioactive components, the largest contribution originates from the flexible PCBs of the MDM and is expected to be dominated by intrinsic contamination of the Kapton substrate, primarily from the decay chains of ${^{228}\mathrm{Th}}$, ${^{226}\mathrm{Ra}}$, and $^{210}\mathrm{Pb}$.
Further contributions arise from intrinsic contaminations of the \cawo\ target crystals~\cite{Muenster:2014cawo4_radiopurity}, the brass screws inside the MDM, the detector-cage PCBs used for the COV readout, the black coating of the radiation shields attached to the detector cage, the cryostat vessels, and the external lead and polyethylene shielding, as listed in Table~\ref{tab:mdm_material_screening}.

In the energy range from \(0\) to \SI{10}{\kilo\electronvolt} explored by the Technical Run, radioactive contamination from the setup components contributes \mbox{\SI{27.4 \pm 0.4}{counts/(kg\,keV\,day)}} to the spectra collected with the \cawo\ target detectors, accounting for approximately \SI{20}{\percent} of the total predicted background of \mbox{\SI{137.2 \pm 1.5}{counts/(kg\,keV\,day)}}, that is expected to be dominated by atmospheric neutrons and ambient gammas.

It is also relevant to compare the expected background levels in the \cevns\ region of interest, between \SI{10}{\electronvolt} and \SI{100}{\electronvolt}, for the Technical Run and for the physics phase. 
The predicted background rate for the Technical Run is approximately five times as high as that of the physics phase~\cite{nucleus_background_paper}. This difference is largely attributable to the absence in the Technical Run configuration of an additional cryogenic neutron shielding foreseen for the physics phase. With this shielding in place and assuming successful LEE suppression, a signal-to-background ratio of approximately \(S/B \gtrsim 1\) in the \cevns\ region of interest is expected to be achievable even with a relatively small target mass as in the MDM setup.

Overall, the simulations indicate that the MDM provides a compact detector interface without constituting a dominant radioactive background source. The Technical Run configuration is therefore suitable for characterizing the remaining background components under reactor-site conditions and for validating the background model before the first physics phase.

\FloatBarrier
\section{Commissioning Results at TUM}
\label{sec:commissioning}

For the Technical Run, the six double-TES \cawo\ detectors introduced in Sec.~\ref{subsec:TES_production_and_characterization} were characterized at TUM, and four of them were prepared for deployment within the MDM at the reactor site. The main goal of these commissioning measurements was to verify that the MDM provides stable operation of the cryogenic target detectors, achieves the performance required for low-threshold reactor-\cevns\ searches, and remains compatible with the surrounding COV.
The primary performance requirement is set by the energy threshold of the \nucleus\ experiment. A threshold of about \SI{20}{eV} is targeted~\cite{nucleus2017}, which corresponds to a baseline resolution of $\sigma_{\mathrm{BL}}=\SI{4}{eV}$ when using the common definition of $E_{\mathrm{thr}} = 5\cdot \sigma_{\mathrm{BL}}$.

The performance of the six target detectors was evaluated in several commissioning runs. Here, one measurement campaign was dedicated to an operating-point optimization based on the signal-to-noise ratio of photon pulses used for calibration~\cite{sensitivityenhancementtechniques}. Another measurement evaluated the simultaneous operation of the MDM and the COV inside the detector cage to assess possible interference between the target detectors and the veto system.
The following subsections describe the analysis procedure, energy calibration, achieved detector performance, and compatibility of the MDM with the COV.

\subsection{Data Processing and Pulse Reconstruction}
\label{sec:data_and_analysis}

Data were recorded with sampling frequencies between \SI{20}{kHz} and \SI{100}{kHz}, representing a compromise between data volume and the time resolution required for pulse reconstruction, in particular for resolving the rising edge of low-energy events and partially saturated high-energy events.

The data were processed using the analysis framework \textsc{cait}, a Python package for cryogenic detector data~\cite{Wagner:2022xde}, adapted for use in \nucleus. Event triggering, selection, and pulse reconstruction follow the approach described in Refs.~\cite{LBR_paper, xrf-paper}. The data stream contains both particle pulses and artificial heater pulses, which are injected to monitor and stabilize detector operation.

Triggering is performed offline using the optimum-filter method~\cite{DiDomizio:2010ph}. For each detector channel, a pulse template and a noise power density are constructed before filtering. The pulse template is obtained by averaging several hundred well-reconstructed particle pulses within the detector's linear response range, where pulse shapes are expected to be comparable. The noise power spectrum is obtained from randomly selected clean baseline traces that do not contain particle events or other artifacts.

The baseline resolution $\sigma_{\mathrm{BL}}$ was evaluated using two approaches. First, a standard one-dimensional optimum filter (1D-OF)~\cite{Gatti1986_optimum_filter} was applied to reconstruct the amplitude of random noise traces of each TES channel individually. Second, a two-dimensional optimum filter (2D-OF)~\cite{sensitivityenhancementtechniques} was used to process both TES channels of one detector simultaneously. Since both TESs measure the same physical event, the combined treatment improves the precision of the amplitude estimator and is expected to reduce the baseline resolution compared to the single-channel case~\cite{sensitivityenhancementtechniques}.
The results are presented in Sec.~\ref{sec:performance_summary}.

\subsection{Pulse-Shape and Noise Characterization}
\label{sec:pulse_template}

\begin{figure}[htbp]
    \centering
    \begin{subfigure}[b]{0.4\columnwidth}
        \centering
        \includegraphics[height=5cm]{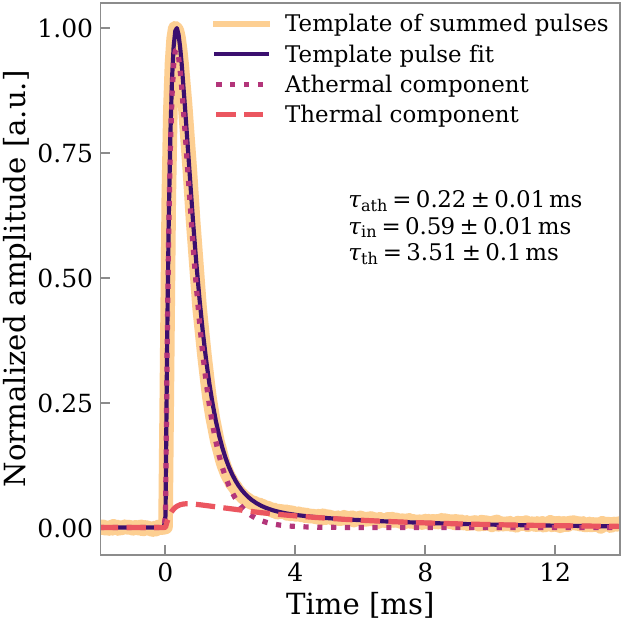}
        \caption{}
        \label{fig:sev_fit}
    \end{subfigure}
    \hfill
    \begin{subfigure}[b]{0.55\columnwidth}
        \centering
        \includegraphics[height=5cm]{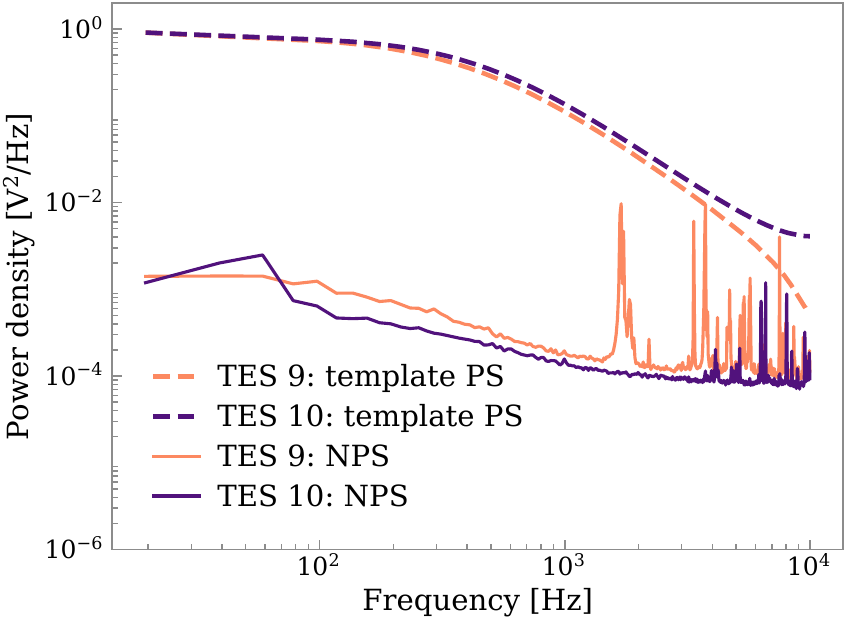}
        \caption{}
        \label{fig:nps}
    \end{subfigure}
    \caption{
    Pulse-shape and noise characterization for detector~D.
    \textbf{(a)} Template pulse obtained by averaging 11 particle events in the linear response range of the detector. The template pulse is fitted with the pulse shape model of Ref.~\cite{Probst:1995hjq}, yielding the characteristic time constants $\tau_{\mathrm{ath}}$, $\tau_{\mathrm{in}}$, and $\tau_{\mathrm{th}}$ with their associated statistical uncertainties. The template pulse line thickness is increased for visualization.
    \textbf{(b)} Noise power spectrum (NPS) and normalized power spectra (PS) of the template pulses for both TES channels. The Fourier-domain pulse templates and the corresponding noise power densities are used to construct the optimum filters for pulse-height reconstruction.
    }
    \label{fig:pulse_and_NPS}
\end{figure}

Fig.~\ref{fig:pulse_and_NPS} shows an example of the pulse template and noise power spectrum for detector~D. The template pulse is fitted using the pulse-shape model developed for cryogenic TES detectors in Ref.~\cite{Probst:1995hjq}. In this model, the absorber is coupled to a TES that is weakly connected to the heat bath, provided here by the cryostat. Following a particle interaction, athermal phonons are generated in the absorber and are partially collected by the TES before the detector relaxes back to equilibrium. The model distinguishes between calorimetric and bolometric operation. 
The \nucleus\ TES design (see Sec.~\ref{subsec:TES_design}) follows the calorimetric approach, motivated by the successful performance of \cresst\ detectors~\cite{Angloher2026CRESST} and optimized for efficient collection of the athermal phonon population. Accordingly, the thermal coupling of the TES to the heat bath is chosen sufficiently weak that the TES relaxation is slow compared with the phonon-collection timescale, resulting in a pulse decay time that is typically about one order of magnitude longer than the rise time.

The measured pulse shape is described as the superposition of an athermal and a thermal signal component, governed by the heat capacities and thermal conductances associated with the couplings between the absorber, TES, and heat bath. Assuming calorimetric operation, the time constant $\tau_{\mathrm{ath}}$ describes the athermal phonon collection time, $\tau_{\mathrm{in}}$ the relaxation time of the TES system through its thermal link to the heat bath, and $\tau_{\mathrm{th}}$ the relaxation time of the absorber crystal after the non-thermal phonon energy input ended. 
Here, the athermal phonon signal is collected before the TES and absorber system return to equilibrium with the heat bath, meaning that the pulse height is proportional to the integrated athermal energy input. In the bolometric regime, by contrast, the TES equilibrates before the athermal energy input has decayed; therefore, the pulse height is proportional to the flux of athermal phonons.
In previous measurements with \nucleus\ detectors featuring different TES designs and absorber configurations, the extracted time constants indicated operation in the calorimetric regime~\cite{LBR_paper,xrf-paper}.

An example template pulse and its pulse-shape fit are presented in Fig.~\ref{fig:sev_fit}.
For the six MDM detectors, comparable pulse shapes are obtained among the detectors, quantified by the mean and the standard deviation of the extracted time constants under the assumption of operating in the calorimetric regime:
\begin{equation*}
	\overline{\tau}_{\mathrm{ath}}=\SI{0.20 \pm 0.04}{ms},\qquad
	\overline{\tau}_{\mathrm{in}}=\SI{0.58 \pm 0.13}{ms},\qquad
	\overline{\tau}_{\mathrm{th}}=\SI{3.66 \pm 0.67}{ms}.
\end{equation*}
As can be seen, $\tau_{\mathrm{ath}}$ and $\tau_{\mathrm{in}}$ are of the same order of magnitude, indicating that the TES relaxation may limit the collection of the athermal phonon signal. Nevertheless, as summarized in Sec.~\ref{sec:performance_summary}, the MDM detectors demonstrated excellent overall performance. In future design iterations, the TES relaxation time could be increased by reducing the thermal coupling between the TES and the heat bath. This may increase the amount of measured phonons and thereby further enhance the detector sensitivity. Such efforts to change the TES geometry are currently being investigated.

Compared with the $5\times5\times5~\si{mm^3}$ \cawo\ \nucleus\ detector presented in Ref.~\cite{LBR_paper}, the pulse time scales of the present detectors are considerably shorter. The athermal rise time $\tau_{\mathrm{ath}}$ is reduced by a factor of~2, while the decay time constants $\tau_{\mathrm{in}}$ and $\tau_{\mathrm{th}}$ are reduced by factors of~24 and~12, respectively. These differences may originate from changes in the TES design. In particular, the phonon collectors of the TES presented in Ref.~\cite{LBR_paper} feature an additional tungsten layer underneath the aluminium film, whereas the phonon collectors of the present design consist only of aluminium. The geometry of the thermal link was also modified while preserving its thermal coupling strength. The removal of the tungsten layer in the present design was motivated by findings from \cresst\ detectors~\cite{Angloher_2026}. The detailed impact of these TES design modifications on the detector pulse shape and performance is currently under further investigation.

Based on the pulse response, the time coincidence window required for operation with the surrounding veto systems at the Chooz reactor site can be estimated. In Ref.~\cite{LBR_paper}, a \SI{240}{\micro\second} coincidence window was used for the \cawo\ detector operated in coincidence with the muon veto and the COV. Owing to the faster pulse response of the present detectors, a correspondingly shorter coincidence window can be used, thereby reducing the accidental veto-induced dead time. Assuming that the required coincidence-window width scales with the detector rise time, a window of about \SI{120}{\micro\second} is expected. For the anticipated muon rate of \SI{325}{\hertz} at the Chooz site, this corresponds to a projected muon-induced dead time of about \SI{4}{\percent}. The actual reduction in veto-induced dead time enabled by the faster detector response will be quantified from dedicated coincidence measurements at the Chooz site.

The noise power spectrum as well as the pulse template in Fourier space are  shown for both TESs of detector~D in Fig.~\ref{fig:nps}. They are used for the optimum filter reconstruction described in Sec.~\ref{sec:data_and_analysis}.

\FloatBarrier
\subsection{Energy Calibration}
\label{sec:energy_cal}

The energy calibration presented here is based on events from the collimated $^{55}$Fe sources integrated into the MDM. The $^{55}$Fe source produces Mn K$_\alpha$ and K$_\beta$ X-ray lines at \SI{5.89}{keV} and \SI{6.49}{keV}, respectively~\cite{NIST_Xray}. In the measurements presented here, these calibration events lie outside the linear response range of the detectors, meaning that the pulse height no longer scales linearly with the energy input. To overcome this issue, a truncated template fit~\cite{trunc_fit_schmaler2010} is applied. In this approach, the template pulse (Fig.~\ref{fig:sev_fit}) is fitted only in the amplitude range that remains within the linear regime of the pulse, while the saturated part of the pulse (during which the detector is outside the linear range) is excluded.

\begin{figure}[htbp]
    \centering
    \begin{subfigure}[b]{0.35\columnwidth}
        \centering
        \includegraphics[height=5cm]{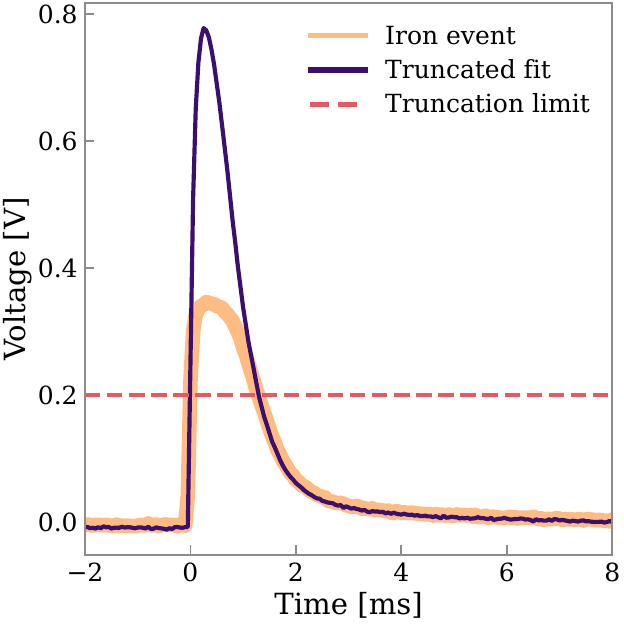}
        \caption{}
        \label{fig:trunc_fit}
    \end{subfigure}
    \hfill
    \begin{subfigure}[b]{0.6\columnwidth}
        \centering
        \includegraphics[height=5cm]{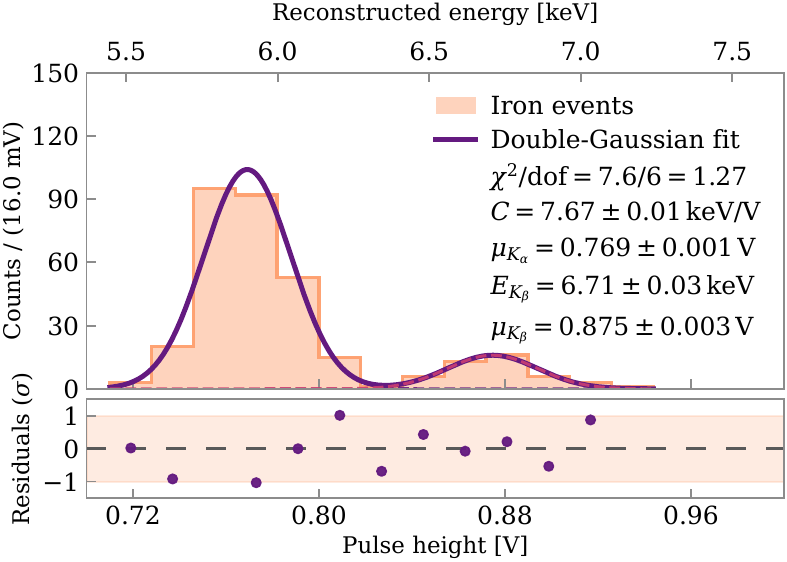}
        \caption{}
        \label{fig:iron_peak}
    \end{subfigure}
    \caption{
    $^{55}$Fe calibration example for detector~D.
    \textbf{(a)} Example of a truncated template fit to a partially saturated $^{55}$Fe pulse. The truncation limit defines the pulse region included in the fit.
    \textbf{(b)} Double-Gaussian fit to the Mn $K_{\alpha}$ and $K_{\beta}$ lines reconstructed with the truncated template fit. The amplitudes, means, and widths of both Gaussian components are treated as free parameters. The calibration factor $C$ is obtained from the fitted $K_{\alpha}$ mean and the associated uncertainty is only statistical. The lower panel shows the residuals in units of the statistical uncertainty of each bin, where the orange area indicates the $1\sigma$ band. The dashed line indicates zero.
    }
    \label{fig:iron_calibration}
\end{figure}

Fig.~\ref{fig:iron_calibration} shows an example of the truncated template fit and the corresponding reconstructed $^{55}$Fe spectrum for detector~D. The voltage at which the pulse shape starts to deviate from the template defines the truncation limit; this is quantified by an increase in the RMS of the template fit, in which the truncation is not yet accounted for, and which indicates that the detector's response becomes increasingly non-linear. In the example of Fig.~\ref{fig:trunc_fit}, the value marking the RMS fit increase is \SI{0.2}{V}, which was set as the truncation limit. The reconstructed pulse-height spectrum is fitted with a double-Gaussian model describing the Mn K$_\alpha$ and K$_\beta$ lines. The calibration factor is determined from the Mn K$_\alpha$ peak position by assuming a linear extrapolation to zero energy.
For the example shown in Fig.~\ref{fig:iron_peak}, the Mn K$_\beta$ line is reconstructed at $E_{\mathrm{K}_{\beta}}^{\mathrm{meas}} = (6.71 \pm 0.03)\,\mathrm{keV}$, which is above the literature value of \SI{6.49}{keV}~\cite{NIST_Xray},
corresponding to a \SI{3}{\percent} deviation of the reconstructed K$_\beta$ energy from the literature value. This deviation may arise from limitations of the truncated template-fit reconstruction, from nonlinear detector response, or from the assumption of a linear extrapolation between the calibration line and zero energy. Sec.~\ref{sec:sys_uncertainties_energy_cal} comments on the systematic uncertainties associated with the X-ray energy calibration.

\FloatBarrier
\subsection{Detector Performance}
\label{sec:performance_summary}

\begin{figure}[htbp]
\centering
\includegraphics[height=7cm]{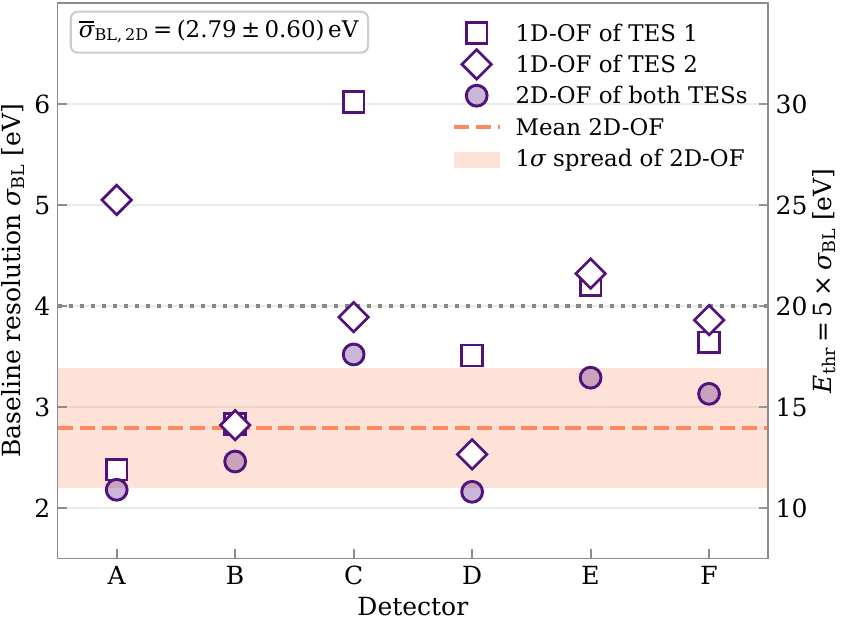}
\caption{
Baseline resolution $\sigma_{\mathrm{BL}}$ and corresponding energy-threshold estimate $E_{\mathrm{thr}}=5\cdot\sigma_{\mathrm{BL}}$ for the best-performing configuration of each \cawo\ detector A--F. The 1D-OF results are shown separately for TES~1 (squares) and TES~2 (diamonds), while the 2D-OF results combining both TES channels are shown as circles. The horizontal dashed line and shaded band indicate the mean 2D-OF baseline resolution and the corresponding standard deviation across the six detectors, respectively. Individual statistical uncertainties are omitted from the figure for clarity and are given in Table~\ref{tab:selected_detector_performance}. The gray dotted line marks $\sigma_{\mathrm{BL}}=\SI{4}{eV}$, corresponding to the targeted energy threshold of $E_{\mathrm{thr}}=\SI{20}{eV}$.
}
\label{fig:br_summary}
\end{figure}

The six \cawo\ double-TES detectors were characterized using the $^{55}$Fe calibration procedure described above. Stable operation was demonstrated for all detectors over measurement periods of up to 21 days. 
Stability at the \SI{10}{\percent}-level or better across all detectors was observed using test pulses that monitored the TES response.
Table~\ref{tab:selected_detector_performance} and Fig.~\ref{fig:br_summary} summarize the baseline resolutions obtained for each detector in its best-performing configuration, together with the corresponding bias currents. Results are presented for the one-dimensional optimum filter (1D-OF), applied separately to the two TES channels, and for the two-dimensional optimum filter (2D-OF)~\cite{sensitivityenhancementtechniques}, which combines the information from both TES channels of a double-TES detector. Table~\ref{tab:selected_detector_performance} additionally lists the relative improvement achieved with the 2D-OF compared with the respective 1D-OF result of each TES channel. Fig.~\ref{fig:br_summary} shows the corresponding energy-threshold estimates based on the common definition $E_{\mathrm{thr}}=5\sigma_{\mathrm{BL}}$.
All baseline resolutions are expressed in electronvolts using the $^{55}$Fe calibration. The uncertainties listed in Table~\ref{tab:selected_detector_performance} include only statistical contributions from the X-ray calibration and the baseline-resolution estimation. Systematic uncertainties associated with the energy calibration are discussed in Sec.~\ref{sec:sys_uncertainties_energy_cal}.

\begin{table*}[htbp]
\centering
\caption{
Results for the best-performing configurations of the six \cawo\ double-TES detectors A--F and their corresponding bias currents. The baseline resolutions were obtained using the individual one-dimensional optimum filters and the combined two-dimensional optimum filter. The relative improvements of the 2D-OF results are given with respect to the 1D-OF results of TES~1 and TES~2, respectively.
}
\label{tab:selected_detector_performance}
\footnotesize
\setlength{\tabcolsep}{3.5pt}
\renewcommand{\arraystretch}{1.}

\begin{tabular}{
    c
    c
    c
    S[table-format=1.2(2)]
    S[table-format=1.2(2)]
    S[table-format=1.2(2)]
    c
}
    \toprule
    &
    \multicolumn{1}{c}{TES channels} &
    \multicolumn{1}{c}{Bias currents} &
    \multicolumn{3}{c}{Baseline resolution} &
    \multicolumn{1}{c}{2D-OF improvement} \\
    \cmidrule(lr){2-2}
    \cmidrule(lr){3-3}
    \cmidrule(lr){4-6}
    \cmidrule(lr){7-7}

    Detector &
    {TES 1 / TES 2} &
    {$I_{\mathrm{b},1}/I_{\mathrm{b},2}$ [\si{\micro\ampere}]} &
    {$\sigma_{\mathrm{BL},1}$ [\si{\electronvolt}]} &
    {$\sigma_{\mathrm{BL},2}$ [\si{\electronvolt}]} &
    {$\sigma_{\mathrm{BL,2D}}$ [\si{\electronvolt}]} &
    {vs.\ TES 1 / TES 2 [\si{\percent}]} \\
    \midrule

    A & 3 / 4
      & {3.0 / 4.0}
      & 2.38(2) & 5.05(1) & 2.18(2)
      & 8.4 / 56.8 \\

    B & 5 / 6
      & {2.0 / 3.0}
      & 2.83(2) & 2.82(2) & 2.46(3)
      & 13.1 / 12.8 \\

    C & 7 / 8
      & {2.5 / 3.0}
      & 6.02(1) & 3.89(1) & 3.52(2)
      & 41.5 / 9.5 \\

    D & 9 / 10
      & {2.5 / 2.5}
      & 3.51(2) & 2.53(2) & 2.16(2)
      & 38.5 / 14.6 \\

    E & 11 / 12
      & {4.5 / 4.5}
      & 3.64(1) & 3.86(1) & 3.13(2)
      & 14.0 / 18.9 \\

    F & 15 / 16
      & {3.5 / 4.0}
      & 4.21(2) & 4.32(3) & 3.29(3)
      & 21.9 / 23.8 \\

    \bottomrule
\end{tabular}
\end{table*}

In general, all individual TESs exhibit excellent and homogeneous baseline resolutions, comparable to or better than those previously reported for \nucleus\ detectors, such as the detectors presented in Ref.~\cite{LBR_paper}, which were even smaller in mass. The single-TES resolution is particularly relevant for the discrimination of the LEE down to the lowest energies.
As shown in Table~\ref{tab:selected_detector_performance}, the 2D-OF improves the baseline resolution relative to the better individual 1D-OF results of both sensors by an average of approximately \SI{14}{\percent}. Across the six detectors, the mean 2D-OF baseline resolution and its standard deviation is \SI{2.79 \pm 0.6}{eV}, clearly demonstrating that the targeted resolution of \SI{4}{eV} required for reactor-\cevns\ measurements with \nucleus\ is accessible with the developed MDM detectors.

For most configurations, the detector working points were selected manually based on operational experience. Detector~D was instead subjected to the dedicated working-point optimization procedure described in Ref.~\cite{sensitivityenhancementtechniques}. This resulted in the best performance obtained in this study, with a 2D-OF baseline resolution of $\sigma_{\mathrm{BL,Fe}} =\SI{2.16\pm0.02}{eV}$, where the quoted uncertainty is statistical and the energy scale is based on the $^{55}$Fe calibration. 
This is a factor of 2 better than the initial threshold goal and represents a state-of-the-art baseline resolution for a cryogenic \cawo\ detector. It improves upon the performance previously reported for a smaller cryogenic \nucleus\ detector in Ref.~\cite{nucleus2017} by \SI{29}{\percent}, whose results are also based on iron calibration.

For the upcoming Technical Run, detectors~A, B, C, and E, were selected for deployment at Chooz. These detectors were operated together in the latest tested configuration and can therefore be deployed without further detector replacement, additional handling, or recommissioning before installation at the reactor site, while detectors~D and F are retained as backups for future upgrades. Applying the dedicated working-point optimization during the Technical Run is expected to further improve the average performance of the deployed detectors. Since the reactor-\cevns\ recoil spectrum rises steeply toward low energies, further reductions in the baseline resolution directly increase the accessible signal region.

\FloatBarrier   
\subsection{Systematic Uncertainties on the Energy Calibration}
\label{sec:sys_uncertainties_energy_cal}

In the study presented in Ref.~\cite{sensitivityenhancementtechniques}, detector~D was used to validate a signal-to-noise-ratio optimization procedure based on a two-dimensional scan of the bias and heater currents, with the performance of each operating point estimated via the response to a reference LED pulse. For the same dataset as presented above, the LED-based calibration~\cite{DelCastello:2024ehl} yields a baseline resolution of $\sigma_{\mathrm{BL,LED}} =\SI{2.94\pm0.05}{eV}$, where the quoted uncertainty is statistical. By comparison, the $^{55}$Fe-based calibration applied in the present work yields $\sigma_{\mathrm{BL,Fe}} = \SI{2.16\pm0.02}{eV}$. 
This difference corresponds to a \SI{26.5}{\percent} lower baseline resolution for the $^{55}$Fe-based calibration than for the LED-based value. The difference is substantially larger than the statistical uncertainties and therefore illustrates the present systematic uncertainty associated with the conversion of detector amplitudes into an absolute energy scale. 
 
Two effects may contribute to this calibration difference. First, deviations from a linear detector response can introduce an energy-dependent systematic effect. In an example study, the detector response was investigated using several characteristic X-ray fluorescence lines between \SI{600}{eV} and \SI{6}{keV}~\cite{xrf-paper}. Deviations from linearity of up to \SI{18}{\percent} were observed even within the nominally linear response range~\cite{xrf-paper}. Such effects may be further enhanced for the $^{55}$Fe calibration in this work because the pulses are outside the linear response range and must be reconstructed using truncated template fits. This can introduce additional systematic uncertainties, especially for data with a limited number of samples on the rising edge.

Second, LED and X-ray calibrations rely on different energy-deposition processes. Differences of approximately \SI{25}{\percent} between calibrations based on LED pulses and based on reconstructed Cu fluorescence lines have also been observed in an independent measurement~\cite{LBR_paper}. Although both methods ultimately produce electronic excitations in the absorber, their initial energy-deposition mechanisms differ: a $^{55}$Fe event results from the absorption of a single keV X-ray photon and a localized energy deposition, whereas an LED pulse deposits its energy through the absorption of many optical photons throughout the absorber volume. These differences may lead to calibration-dependent reconstructed energies, and a method-dependent systematic uncertainty associated with the LED calibration cannot presently be excluded.

A further limitation is that neither calibration method directly reproduces the nuclear-recoil signal expected from \cevns. Therefore, a precise calibration with well-defined nuclear recoils in the \cevns\ region of interest is particularly relevant for validating the energy scale. Calibration approaches providing sub-keV nuclear recoils have been developed within the CRAB project~\cite{PhysRevLett.130.211802,2025CRAB}. 
For \nucleus, an LED-based calibration will be used during reactor operation, with the LED response anchored to dedicated nuclear-recoil measurements and the detector linearity mapped throughout the region of interest. For the first \nucleus\ physics phase with a \cawo\ target mass of approximately \SI{7}{\gram}, an energy-scale uncertainty of up to \SI{25}{\percent} is still acceptable, as it contributes only at the few-percent level to the total uncertainty and is therefore subdominant to the statistical uncertainty~\cite{2026prospectnucleusexperimentchooz-xzr1}. For future precision measurements with increased exposure, however, improved control of the energy scale will become increasingly important, making this combined calibration strategy a promising approach for reducing the corresponding systematic uncertainty.

\FloatBarrier
\subsection{Integration with the Cryogenic Outer Veto}
\label{sec:mdm_cov_compatibility}

The compatibility of the MDM with the surrounding COV was tested by operating the target detectors simultaneously with the six COV crystals installed inside the detector cage. The purpose of this measurement was to verify stable operation of the integrated detector system and to investigate whether the operation of either detector subsystem affects the performance of the other.

\begin{figure}[htbp]
    \centering
    \begin{subfigure}[b]{0.40\columnwidth}
        \centering
        \includegraphics[height=5cm]
        {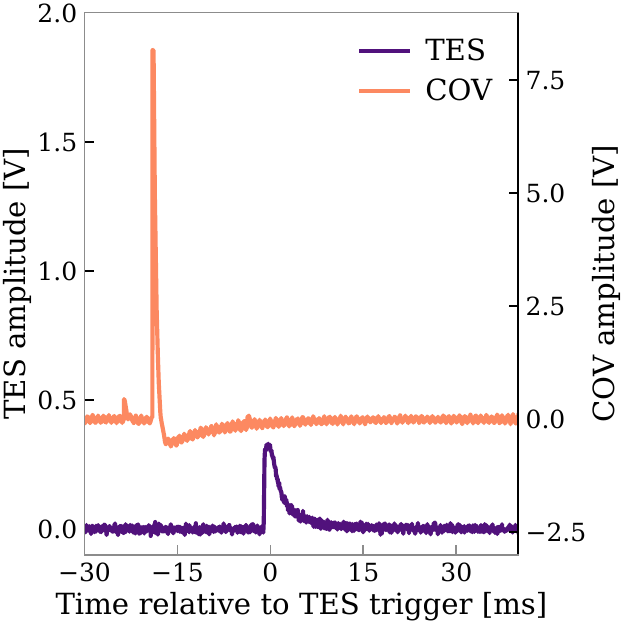}
        \caption{}
        \label{fig:cov_cross_talk}
    \end{subfigure}
    \hfill
    \begin{subfigure}[b]{0.55\columnwidth}
        \centering
        \includegraphics[height=5cm]
        {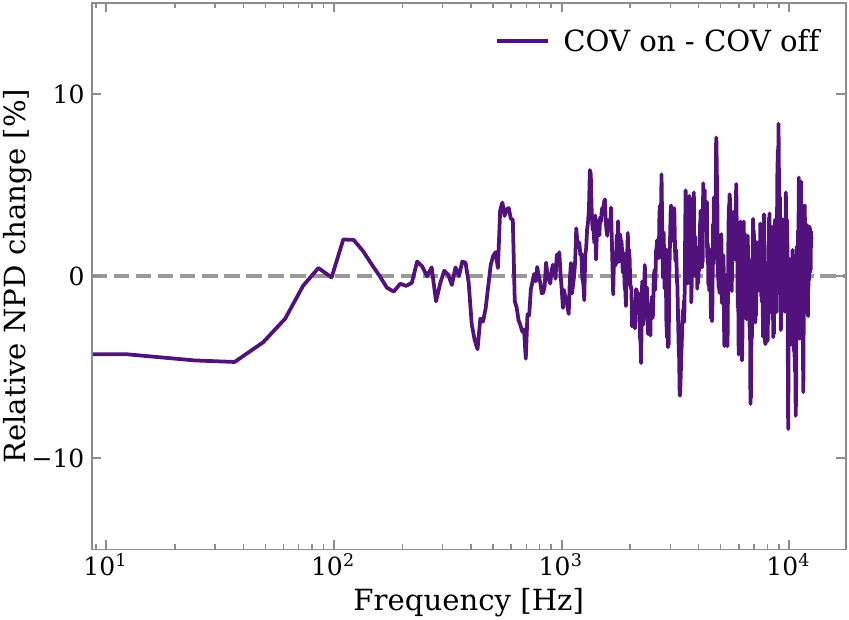}
        \caption{}
        \label{fig:cov_nps}
    \end{subfigure}
    \caption{
    Compatibility test of detector~D operated simultaneously with the
    COV.
    \textbf{(a)} Example of a high-energy event with an energy of $\mathcal{O}(\SI{10}{keV})$ producing a large pulse in one COV crystal without a corresponding signal above the noise level in the TES channel, together with a TES event with an energy of $\mathcal{O}(\SI{1}{keV})$ without a measurable response in the COV channel.
    \textbf{(b)} Relative change in the averaged TES noise power density between data acquired with the COV electrodes switched on and off. The relative difference is defined with respect to the configuration with the COV electrodes switched off, and the dashed line indicates zero.
    }
    \label{fig:cov_compatibility}
\end{figure}

The cross-talk analysis focused on one TES channel of detector~D and one representative COV crystal. Fig.~\ref{fig:cov_cross_talk} shows two example events. A high-energy event with an energy of $\mathcal{O}(\SI{10}{keV})$ producing a large pulse in the COV channel does not generate a corresponding signal above the noise level in the TES channel. Additionally, a pulse observed in the TES channel with an energy of $\mathcal{O}(\SI{1}{keV})$ is not accompanied by a measurable signal in the COV channel. This behavior was observed consistently and demonstrates the simultaneous readout of the two detector systems without observable cross-talk in either direction.

Possible noise coupling from the COV operation into the TES channel was investigated by comparing the averaged noise power densities measured with the COV electrodes switched on and off. The relative change in the TES noise power density is shown in Fig.~\ref{fig:cov_nps}. The relative difference fluctuates around zero and remains within approximately \SI{10}{\percent} over the displayed frequency range. In particular, no systematic increase in the TES noise is observed during operation of the COV electrodes.
This result is further supported by comparing the TES baseline resolutions with the COV electrodes switched on and off. The two measurements yield consistent baseline resolutions, confirming that operation of the COV electrodes does not measurably degrade the TES noise performance.

Conversely, the influence of TES operation on the COV performance was also investigated by comparing the COV response during simultaneous operation of both detector systems. No measurable change in the COV signal or noise performance was observed, indicating that operation of the target-detector TESs is likewise compatible with stable COV operation.

These measurements demonstrate the successful simultaneous operation of the MDM detectors and the COV, thereby validating the electrical-isolation concept described in Sec.~\ref{sec:MDM_design}. The absence of observable cross-talk or noise degradation confirms that both detector systems can be operated independently without mutual interference, demonstrating the compatibility of the integrated configuration for operation at the reactor site.

\FloatBarrier
\section{Conclusion and Outlook}
\label{sec:conclusion}

In this work, the cryogenic target-detector module for the \nucleus\ Technical Run was developed, constructed, and commissioned at the Technical University of Munich. The MDM integrates four gram-scale \cawo\ double-TES detectors with a total target mass of approximately \SI{7}{g} into a compact holder compatible with the limited volume inside the Cryogenic Outer Veto. It provides the mechanical support, thermalization, signal routing, and interchangeable $^{55}$Fe and LED calibration interfaces required for operation within the \nucleus\ cryogenic infrastructure. Geant4-based simulations indicate that radioactive contaminations of the module and its immediate surroundings are not expected to dominate the Technical Run background.

The TESs produced on a common \cawo\ substrate showed a homogeneous transition-temperature distribution of $T_{50}=(14.49\pm0.31)$~\si{\milli\kelvin}, allowing six double-TES detectors with matched transition temperatures to be selected and characterized. Stable operation was demonstrated over periods of up to three weeks. Taking advantage of the double-TES structure, the 2D optimum filter~\cite{sensitivityenhancementtechniques} improved the baseline resolution by approximately \SI{14}{\percent} on average compared with the individual 1D optimum filters. Across all six detectors, a mean energy baseline resolution of $\Bar{\sigma}_{\mathrm{BL}}$=\SI{2.79 \pm 0.6}{eV} was achieved using the $^{55}$Fe-calibration and the two-dimensional optimum filter of the double-TES detectors.
Assuming the commonly used energy threshold definition $E_{\mathrm{thr}}=5\cdot\sigma_{\mathrm{BL}}$, this corresponds to a mean threshold below the \nucleus\ performance goal of \SI{20}{\electronvolt}~\cite{nucleus2017}, and demonstrates that the threshold scale required for reactor-\cevns\ studies with \nucleus\ is achievable.
The best-performing detector reached a baseline energy resolution of $\sigma_{\mathrm{BL}} =(2.16 \pm 0.02_{\mathrm{stat}})$~\si{\electronvolt}, based on $^{55}$Fe calibration, and after implementing dedicated working-point optimizations~\cite{sensitivityenhancementtechniques}. This improves upon the \nucleus\ performance goal by nearly a factor of two and represents the best reported performance of a cryogenic \cawo\ detector.

Simultaneous operation of the target detectors and the COV was also demonstrated. No observable cross-talk between the two detector systems was found, and operation of the COV electrodes produced no systematic increase in the TES noise or baseline resolution. Together with the demonstrated detector stability and performance, these measurements validate the integrated configuration for deployment at the Chooz nuclear power plant. The Technical Run will test the long-term operation of the complete experimental infrastructure under reactor-site conditions and characterize the background composition at the experimental site. Although the Standard Model \cevns\ sensitivity may be limited by the low-energy excess during this run, the achieved thresholds enable competitive searches for several beyond-the-Standard-Model scenarios~\cite{2026prospectnucleusexperimentchooz-xzr1}.

In addition, the measured pulse time scales indicate that the MDM detectors can be operated efficiently in coincidence with the surrounding veto systems at the Chooz reactor site, with an expected muon-induced dead time of only \SI{4}{\percent}.
Taken together, these results demonstrate the readiness of the target-detector system for the \nucleus\ Technical Run at Chooz and constitute a key milestone toward future reactor-\cevns\ measurements with cryogenic detectors.

Further developments will address the main limitations identified during commissioning. The working-point optimization demonstrated for one detector will be extended to the deployed detectors, while modifications of the TES design will be investigated to further improve their sensitivity. 
Although the \cevns\ sensitivity of the first \nucleus\ Physics Run is expected to be dominated by statistical rather than energy-calibration uncertainties, the LED calibration foreseen for reactor operation will be cross-checked by anchoring the energy scale within the region of interest to well-defined sub-keV nuclear recoils developed within the CRAB program~\cite{PhysRevLett.130.211802,2025CRAB}. This will provide an independent validation of the energy reconstruction and constrain the associated systematic uncertainties.
For this subsequent Physics Run, the passive MDM holder is planned to be replaced by an instrumented inner-veto system capable of identifying holder-related and stress-induced events, complementing the rejection of sensor-related events provided by the double-TES readout~\cite{Schermer_Nicole_2025_detectors_proceeding,LBR_paper}. Together, improved detector optimization, a validated low-energy calibration, and dedicated mitigation of the low-energy excess will provide the basis for future reactor-\cevns\ measurements with cryogenic detectors.

\section*{Appendix}
Table~\ref{tab:mdm_material_screening} summarizes the material-screening results for the MDM components and their immediate surroundings. In rare-event searches, radioactive contaminants in the materials of the experimental setup can contribute directly to the background in the \cevns\ region of interest. Relevant isotopes may be of primordial origin, including $^{232}$Th, $^{238}$U, $^{235}$U, $^{40}$K, and their progenies; cosmogenic origin, arising from the activation of materials by cosmic rays; or anthropogenic origin, such as $^{60}$Co and $^{137}$Cs. The materials used in the \nucleus\ setup were therefore characterized by non-destructive gamma-ray spectrometry at the \stella\ facility~\cite{Laubenstein:2017yjj}.

The screening results for the mechanical structures, internal and external shielding components, and cryostat vessels have already been reported in Ref.~\cite{nucleus_background_paper}. Table~\ref{tab:mdm_material_screening} therefore focuses on the components located in or immediately surrounding the detector cage that are relevant for the Technical Run. The value for the high-purity copper, which is also used for the copper structures of the present setup, was taken from Ref.~\cite{Angloher_2024_CRESST_BG}.
A small number of components located close to the cryogenic target detectors or the detector cage have not yet been screened, including the optical fibers used for the LED calibration, the sapphire support spheres, and the COV readout cables. Owing to their small masses and their positions within the experimental setup, these components are not expected to contribute significantly to the background in the \cevns\ region of interest. Dedicated screening measurements are nevertheless planned to verify this assumption and to complete the radiopurity assessment of the detector environment.

\begin{sidewaystable}[p]
    \centering
    \caption{
    Results of the screening measurements of the main components of the cryogenic detector module and its immediate surroundings for the \nucleus\ Technical Run. Specific activities are given in \si{\becquerel\per\kilogram} for components quantified by mass, whereas activities are given in \si{\milli\becquerel\per\piece} for components screened as complete pieces, particularly for components consisting of multiple materials.
    Upper limits are given at the \SI{90}{\percent} confidence level.
    }
    \label{tab:mdm_material_screening}

    \fontsize{5}{5}\selectfont
    \setlength{\tabcolsep}{2.6pt}
    \renewcommand{\arraystretch}{1.15}

    \begin{tabular}{ >{\raggedright\arraybackslash}p{3.cm} ccc rrrrrrr l }
        \toprule

        & 
        & 
        & 
        & \multicolumn{3}{c}{\textbf{$^{238}$U series}}
        & \multicolumn{2}{c}{\textbf{$^{232}$Th series}}
        & 
        & 
        & 
        \\

        \cmidrule(lr){5-7}
        \cmidrule(lr){8-9}

        \textbf{Component}
        & \textbf{Mass [kg]}
        & \textbf{Amount}
        & \textbf{Material}
        & \textbf{$^{238}$U}
        & \textbf{$^{226}$Ra}
        & \textbf{$^{210}$Pb}
        & \textbf{$^{228}$Ra}
        & \textbf{$^{228}$Th}
        & \textbf{$^{40}$K}
        & \textbf{$^{235}$U}
        & \textbf{Unit}
        \\
        \midrule

        Detector cage
			& 2.2
			& -
			& Copper
			& $<7.60{\times}10^{-4}$
			& $<2.00{\times}10^{-5}$
			& -
			& $<2.40{\times}10^{-5}$
			& $<2.00{\times}10^{-5}$
			& $<1.90{\times}10^{-4}$
			& $<5.00{\times}10^{-5}$
			& \si{\becquerel\per\kilogram}
			\\
			
			MDM holder
			& 0.2
			& -
			& Copper
			& $<7.60{\times}10^{-4}$
			& $<2.00{\times}10^{-5}$
			& -
			& $<2.40{\times}10^{-5}$
			& $<2.00{\times}10^{-5}$
			& $<1.90{\times}10^{-4}$
			& $<5.00{\times}10^{-5}$
			& \si{\becquerel\per\kilogram}
			\\
			
			\makecell[l]{Black paint on\\detector-cage\\radiation shields}
			& 0.025
			& -
			& Acrylic
			& 7.00
			& 5.3
			& -
			& 6.2
			& 5.9
			& 4.9
			& 0.44
			& \si{\becquerel\per\kilogram}
			\\
			
			Bolts and nuts
			& 0.25
			& -
			& Brass
			& $<0.042$
			& 0.021
			& - 
			& $<0.0091$
			& 0.008
			& $<0.046$
			& $<0.0025$
			& \si{\becquerel\per\kilogram}
			\\
			
			Detector cage PCB
			& -
			& 1
			& \makecell[l]{Fiberglass\\laminate}
			& 220
			& 29
			& - 
			& 18
			& 46
			& 154
			& 18
			& \si{\milli\becquerel\per pc}
			\\
			
			Detector cage COV connector
			& -
			& 1
			& Nylon 46
			& 21
			& 22
			& 21
			& 0.032
			& 0.047
			& 54
			& 1.2
			& \si{\milli\becquerel\per pc}
			\\
			
			MDM connectors
			& -
			& 2
			& Nylon 46
			& $<5.6$
			& 0.5
			& -
			& $<0.44$
			& $<0.28$
			& $<3.0$
			& $<0.2$
			& \si{\milli\becquerel\per pc}
			\\
			
			MDM support spring
			& 0.0014
			& -
			& Bronze
			& $<0.094$
			& $<0.0040$
			& $<1.5$
			& $<0.0059$
			& $<0.0080$
			& $<0.044$
			& $<0.0052$
			& \si{\becquerel\per\kilogram}
			\\
			
			MDM support structures
			& 0.00044
			& -
			& PTFE
			& $<0.26$
			& $<0.047$
			& $<0.48$
			& $<0.099$
			& $<0.058$
			& $<0.62$
			& $<0.018$
			& \si{\becquerel\per \kilogram}
			\\
			
			\makecell[l]{MDM flex PCB\\inside holder}
			& -
			& 2
			& \makecell[l]{Kapton--\\Copper}
			& $<0.66$
			& 0.0328
			& -
			& $<0.044$
			& 0.03
			& $<0.39$
			& $<0.026$
			& \si{\milli\becquerel\per pc}
			\\
			
			\makecell[l]{MDM flex PCB\\outside holder with\\additional Kapton layer}
			& -
			& 2
			& \makecell[l]{Kapton--\\Copper}
			& $<2.9$
			& 0.22
			& -
			& 0.25
			& 0.25
			& $<1.5$
			& $<0.069$
			& \si{\milli\becquerel\per pc}
			\\
			
			MDM flex PCB glue
			& 0.0003
			& -
			& Epoxy
			& $<13$
			& $<0.78$
			& -
			& $<0.75$
			& $<0.51$
			& $<5.9$
			& $<0.55$
			& \si{\becquerel\per\kilogram}
			\\
			
			Target detector clamps
			& 0.00023
			& -
			& Bronze
			& $<0.094$
			& $<0.004$
			& $<1.5$
			& $<0.0059$
			& $<0.0080$
			& $<0.044$
			& $<0.0052$
			& \si{\becquerel\per\kilogram}
			\\
			
			Target detector material
			& 0.00696
			& -
			& CaWO$_4$
			& $3.10{\times}10^{-3}$
			& $1.74{\times}10^{-4}$
			& $1.08{\times}10^{-4}$
			& $3.30{\times}10^{-5}$
			& $2.40{\times}10^{-5}$
			& $<1.00{\times}10^{-5}$
			& $3.00{\times}10^{-4}$
			& \si{\becquerel\per\kilogram}
        \\

        \bottomrule
    \end{tabular}
\end{sidewaystable}

\section*{Acknowledgements}
This work has been financed by the CEA, the INFN, the ÖAW and partially supported by the TU Munich and the MPI für Physik. \nucleus\ members acknowledge additional funding by the DFG through the SFB1258 and the Excellence Cluster ORIGINS, by the European Commission through the ERC-StG2018-804228 ``NU-CLEUS'', by the P2IO LabEx (ANR-10-LABX-0038) in the framework ``Investissements d'Avenir'' (ANR-11-IDEX-0003-01) managed by the Agence Nationale de la Recherche (ANR), France, by the Austrian Science Fund (FWF) through the ``P 34778-N, ELOISE'', and by Max-Planck-Institut  für Kernphysik (MPIK), Germany. This research was funded in whole or in part by the Austrian Science Fund (FWF) I6955 DOI 10.55776/I6955.
The \nucleus\ collaboration thanks the LNGS staff running the \stella\ facility for the material screening campaigns.

\FloatBarrier
\section*{Data Availability}
The data that support the findings of this article are not
publicly available. The data are available from the authors
upon reasonable request.

\bibliography{MDM_paper_bib}

@article{nucleus2019,
    author = "Angloher, G. and others",
    collaboration = "NUCLEUS",
    title = "{Exploring $\hbox {CE}\nu \hbox {NS}$ with NUCLEUS at the Chooz nuclear power plant}",
    eprint = "1905.10258",
    archivePrefix = "arXiv",
    primaryClass = "physics.ins-det",
    doi = "10.1140/epjc/s10052-019-7454-4",
    journal = "Eur. Phys. J. C",
    volume = "79",
    number = "12",
    pages = "1018",
    year = "2019"
}

@article{strauss2017,
    author = "Strauss, R. and others",
    collaboration = "NUCLEUS",
    title = "{Gram-scale cryogenic calorimeters for rare-event searches}",
    eprint = "1704.04317",
    archivePrefix = "arXiv",
    primaryClass = "physics.ins-det",
    doi = "10.1103/PhysRevD.96.022009",
    journal = "Phys. Rev. D",
    volume = "96",
    number = "2",
    pages = "022009",
    year = "2017"
}

@article{nucleus2017,
    author = "Strauss, R. and others",
    title = "{The $\nu$-cleus experiment: A gram-scale fiducial-volume cryogenic detector for the first detection of coherent neutrino-nucleus scattering}",
    eprint = "1704.04320",
    archivePrefix = "arXiv",
    primaryClass = "physics.ins-det",
    doi = "10.1140/epjc/s10052-017-5068-2",
    journal = "Eur. Phys. J. C",
    volume = "77",
    pages = "506",
    year = "2017"
}

@Article{LBR_paper,
  author        = {Abele, H. and others},
  title         = "{Commissioning of the NUCLEUS Experiment at the Technical University of Munich}",
  doi           = {10.1103/c95p-8kh2},
  issue         = {7},
  pages         = {072013},
  url           = {https://link.aps.org/doi/10.1103/c95p-8kh2},
  volume        = {112},
  collaboration = {NUCLEUS Collaboration},
  journal       = {Phys. Rev. D},
  month         = {Oct},
  numpages      = {21},
  publisher     = {American Physical Society},
  year          = {2025},
}

@misc{nucleus_background_paper,
      title="{Particle background characterization and prediction for the NUCLEUS reactor CE$\nu$NS experiment}", 
      author={H. Abele and others},
      year={2025},
      eprint={2509.03559},
      archivePrefix={arXiv},
      primaryClass={physics.ins-det},
      url={https://arxiv.org/abs/2509.03559}, 
}

@misc{sensitivityenhancementtechniques,
      title="{Sensitivity enhancement techniques for cryogenic calorimeters in the NUCLEUS experiment}", 
      author={M. Cappelli and A. Wallach and others},
      year={2026},
      eprint={2603.28276},
      archivePrefix={arXiv},
      primaryClass={physics.ins-det},
      url={https://arxiv.org/abs/2603.28276}, 
}

@article{2026prospectnucleusexperimentchooz-xzr1,
  title = {{Prospect of the NUCLEUS experiment at Chooz for coherent elastic neutrino-nucleus scattering and new physics searches}},
  author = {Abele, H. and others},
  collaboration = {NUCLEUS Collaboration},
  journal = {Phys. Rev. D},
  volume = {114},
  issue = {1},
  pages = {012016},
  numpages = {20},
  year = {2026},
  month = {Jul},
  publisher = {American Physical Society},
  doi = {10.1103/dnyj-xzr1},
  url = {https://link.aps.org/doi/10.1103/dnyj-xzr1}
}

@article{Baxter_2025,
   title="{Low-Energy Backgrounds in Solid-State Phonon and Charge Detectors}",
   volume={75},
   ISSN={1545-4134},
   url={http://dx.doi.org/10.1146/annurev-nucl-121423-100849},
   DOI={10.1146/annurev-nucl-121423-100849},
   number={1},
   journal={Annual Review of Nuclear and Particle Science},
   publisher={Annual Reviews},
   author={Baxter, Daniel and Essig, Rouven and Hochberg, Yonit and Kaznacheeva, Margarita and von Krosigk, Belina and Reindl, Florian and Romani, Roger K. and Wagner, Felix},
   year={2025},
   month={Sept}, 
   pages={301–326} 
}

@Article{Schermer_Nicole_2025_detectors_proceeding,
  author         = {Schermer, Nicole},
  title          = "{Development of the NUCLEUS Detector to Explore Coherent Elastic Neutrino-Nucleus Scattering}",
  doi            = {10.3390/particles8010008},
  issn           = {2571-712X},
  number         = {1},
  url            = {https://www.mdpi.com/2571-712X/8/1/8},
  volume         = {8},
  article-number = {8},
  journal        = {Particles},
  year           = {2025},
}

@article{DiDomizio:2010ph,
    author = "Di Domizio, S. and Orio, F. and Vignati, M.",
    title = "{Lowering the energy threshold of large-mass bolometric detectors}",
    eprint = "1012.1263",
    archivePrefix = "arXiv",
    primaryClass = "astro-ph.IM",
    doi = "10.1088/1748-0221/6/02/P02007",
    journal = "J. Instrum.",
    volume = "6",
    pages = "P02007",
    year = "2011"
}

@article{Wex:2025jwu,
    author = "Wex, A. and others",
    title = "{Decoupling pulse tube vibrations from a dry dilution refrigerator at milli-Kelvin temperatures}",
    eprint = "2501.04471",
    archivePrefix = "arXiv",
    primaryClass = "physics.ins-det",
    doi = "10.1088/1748-0221/20/05/P05022",
    journal = "J. Instrum.",
    volume = "20",
    number = "05",
    pages = "P05022",
    year = "2025"
}

@misc{blueforsCryostat,
  author    = {{Bluefors Oy}},
  title     = "{LD Dilution Refrigerator Measurement System}",
  url       = {https://bluefors.com/products/dilution-refrigerator-measurement-systems/ld-dilution-refrigerator-measurement-system/},
  access = "05.06.2026"
}

@article{xrf-paper,
    author = "Abele, H. and others",
    title = "{Sub-keV Electron Recoil Calibration for Cryogenic Detectors using a Novel X-ray Fluorescence Source}",
    url={http://dx.doi.org/10.1007/s10909-025-03330-2},
    doi={10.1007/s10909-025-03330-2},
    publisher={Springer Science and Business Media LLC},
    journal={J. Low Temp. Phys.},
    eprint = "2505.17686",
    archivePrefix = "arXiv",
    primaryClass = "physics.ins-det",
    month = "9",
    year = "2025"
}

@article{PhysRevLett.130.211802,
  title = "{Observation of a Nuclear Recoil Peak at the 100 eV Scale Induced by Neutron Capture}",
  author = {Abele, H. and others},
  collaboration = {CRAB Collaboration and NUCLEUS Collaboration},
  journal = {Phys. Rev. Lett.},
  volume = {130},
  issue = {21},
  pages = {211802},
  numpages = {6},
  year = {2023},
  month = {May},
  publisher = {American Physical Society},
  doi = {10.1103/PhysRevLett.130.211802},
  url = {https://link.aps.org/doi/10.1103/PhysRevLett.130.211802}
}

@article{2025CRAB,
  author  = {Abele, H. and others},
  title   = "{The CRAB facility at the TU Wien TRIGA reactor: status and related physics program}",
  journal = {The European Physical Journal C},
  year    = {2025},
  volume  = {85},
  number  = {10},
  pages   = {1188},
  doi     = {10.1140/epjc/s10052-025-14809-3},
  url     = {https://doi.org/10.1140/epjc/s10052-025-14809-3}
}

@article{Probst:1995hjq,
	author = {Pr\"obst, F. and Frank, M. and Cooper, S. and Colling, P. and Dummer, D. and Ferger, P. and Forster, G. and Nucciotti, A. and Seidel, W. and Stodolsky, L.},
	title = "{Model for cryogenic particle detectors with superconducting phase transition thermometers}",
	doi = "10.1007/BF00753837",
	journal = "J. Low Temp. Phys.",
	volume = "100",
	number = "1",
	pages = "69--104",
	year = "1995"
}

@phdthesis{goupyPhd,
    author = "Goupy, C.",
    title = "{Background mitigation strategy for the detection of coherent elastic scattering of reactor antineutrinos on nuclei with the NUCLEUS experiment}",
    school = "Universit\'e Paris Cit\'e",
    year = "2024",
    url = "https://theses.fr/s310276?domaine=theses"
}

@phdthesis{wexPhd,
    author = "Wex, A.",
    title = "{Background Suppression and Cryogenic Vibration Decoupling for the Coherent Elastic Neutrino Scattering Experiment NUCLEUS}",
    school = "Technische Universität München",
    year = "2025",
    url = "https://mediatum.ub.tum.de/?id=1775381"
}

@article{Laubenstein:2017yjj,
    author = "Laubenstein, M. and others",
    title = "{Screening of materials with high purity germanium detectors at the Laboratori Nazionali del Gran Sasso}",
    doi = "10.1142/S0217751X17430023",
    journal = "Int. J. Mod. Phys. A",
    volume = "32",
    number = "30",
    pages = "1743002",
    year = "2017"
}

@phdthesis{rothePhd,
	author = {Rothe, J. F. M. },
	title = "{Low-Threshold Cryogenic Detectors for Low-Mass Dark Matter Search and Coherent Neutrino Scattering}",
	year = {2021},
	school = {Technische Universität München},
	url = {https://mediatum.ub.tum.de/1576351},
}

@unpublished{SchermerThesisPrep2026,
  author = {Nicole Schermer},
  title  = {Development and Commissioning of the Cryogenic Detectors for the NUCLEUS Experiment},
  note   = {PhD thesis, Technical University of Munich, in preparation},
  year   = {2026}
}

@PhdThesis{ErhartPhD,
  author   = {Erhart, Andreas},
  title    = {Commissioning and Calibration of the NUCLEUS Experiment},
  language = {en},
  pages    = {207},
  url      = {https://mediatum.ub.tum.de/1834141},
  school   = {Technische Universität München},
  year     = {2025},
}

@unpublished{vdaq3,
    author       = {{CRYOCLUSTER}},
    howpublished = {Internal communication},
    title = "{The Versatile Data Acquisition}",
    note = {(unpublished)},
    year = {2026}
}

@misc{MultiCB_Flex,
  author       = {{Multi Circuit Boards}},
  title        = "{Flexible Circuit Boards}",
  year         = {2026},
  url          = {https://www.multi-circuit-boards.eu/produkte/leiterplatten/flexible.html},
  note         = {Accessed: 2026-04-27}
}

@misc{Goodfellow_CV00_FL_000130,
  author       = {{Goodfellow Cambridge Ltd}},
  title        = "{Copper (OFHC) Foil, 99.95\% Purity, Part No. CV00-FL-000130}",
  year         = {2026},
  url          = {https://www.goodfellow.co.kr/en/product/copper-o-f-h-c-foil-CV00-FL-000130.htm},
  note         = {Accessed: 2026-04-27}
}

@misc{StarCryo_SQ100,
  author       = {{STAR Cryoelectronics}},
  title        = "{LTS Sensors -- SQ100 Low-Tc DC SQUID Sensor}",
  year         = {2026},
  url          = {https://starcryo.com/lts-sensors/},
  urldate      = {2026-06-06}
}

@article{Angloher_2024_CRESST_BG,
   title="{High-dimensional Bayesian likelihood normalisation for CRESST’s background model}",
   volume={19},
   ISSN={1748-0221},
   url={http://dx.doi.org/10.1088/1748-0221/19/11/P11013},
   DOI={10.1088/1748-0221/19/11/p11013},
   number={11},
   journal={Journal of Instrumentation},
   publisher={IOP Publishing},
   author={Angloher, G. and others},
   year={2024},
   month=Nov, pages={P11013} 
}

@article{Muenster:2014cawo4_radiopurity,
    author        = {M{\"u}nster, A. and others},
    title         = "{Radiopurity of {CaWO$_4$} Crystals for Direct Dark Matter Search with {CRESST} and {EURECA}}",
    journal       = {Journal of Cosmology and Astroparticle Physics},
    volume        = {2014},
    number        = {05},
    pages         = {018},
    year          = {2014},
    doi           = {10.1088/1475-7516/2014/05/018},
    eprint        = {1403.5114},
    archivePrefix = {arXiv},
    primaryClass  = {physics.ins-det}
}

@misc{NIST_Xray,
  author       = {{National Institute of Standards and Technology (NIST)}},
  title        = "{X-ray Transition Energies Database}",
  year         = {2024},
  url          = {https://physics.nist.gov/PhysRefData/XrayTrans/Html/search.html},
  note         = {Accessed: 2026-04-24}
}

@book{Lassner1999Tungsten,
  author    = {Lassner, Erik and Schubert, Wolf-Dieter},
  title     = "{Tungsten: Properties, Chemistry, Technology of the Element, Alloys, and Chemical Compounds}",
  publisher = {Springer},
  address   = {New York},
  year      = {1999},
  edition   = {1},
  isbn      = {978-0-306-45053-2},
  doi       = {10.1007/978-1-4615-4907-9},
  url       = {https://doi.org/10.1007/978-1-4615-4907-9}
}

@article{Angloher2016,
  author  = {Angloher, G. and Bauer, P. and Ferreiro, N. and Hauff, D. and Tanzke, A. and Strauss, R. and Kiefer, M. and Petricca, F. and Reindl, F. and Seidel, W. and Pröbst, F. and Wüstrich, M.},
  title   = {Quasiparticle Diffusion in {CRESST} Light Detectors},
  journal = {Journal of Low Temperature Physics},
  year    = {2016},
  volume  = {184},
  number  = {1--2},
  pages   = {323--329},
  doi     = {10.1007/s10909-016-1512-1}
}

@article{RevModPhys.26.277,
  title = "{Superconducting Elements}",
  author = {Eisenstein, Julian and others},
  journal = {Rev. Mod. Phys.},
  volume = {26},
  issue = {3},
  pages = {277--291},
  numpages = {0},
  year = {1954},
  month = {Jul},
  publisher = {American Physical Society},
  doi = {10.1103/RevModPhys.26.277},
  url = {https://link.aps.org/doi/10.1103/RevModPhys.26.277}
}

@article{Gatti1986_optimum_filter,
  author       = {E. Gatti and P. F. Manfredi},
  title        = "{Processing the signals from solid-state detectors in elementary-particle physics}",
  journal      = {La Rivista del Nuovo Cimento},
  year         = {1986},
  volume       = {9},
  number       = {1},
  pages        = {1--146},
  doi          = {10.1007/BF02822156},
  url          = {https://doi.org/10.1007/BF02822156}
}

@article{Freedman:1973yd,
    author = "Freedman, D. Z.",
    title = "{Coherent Neutrino Nucleus Scattering as a Probe of the Weak Neutral Current}",
    reportNumber = "NAL-PUB-73-76-THY, FERMILAB-PUB-73-076-T",
    doi = "10.1103/PhysRevD.9.1389",
    journal = "Phys. Rev. D",
    volume = "9",
    pages = "1389--1392",
    year = "1974"
}

@article{COHERENT:2017ipa,
    author = "Akimov, D. and others",
    collaboration = "COHERENT",
    title = "{Observation of Coherent Elastic Neutrino-Nucleus Scattering}",
    eprint = "1708.01294",
    archivePrefix = "arXiv",
    primaryClass = "nucl-ex",
    doi = "10.1126/science.aao0990",
    journal = "Science",
    volume = "357",
    number = "6356",
    pages = "1123--1126",
    year = "2017"
}

@article{COHERENT:2020iec,
    author = "Akimov, D. and others",
    collaboration = "COHERENT",
    title = "{First Measurement of Coherent Elastic Neutrino-Nucleus Scattering on Argon}",
    eprint = "2003.10630",
    archivePrefix = "arXiv",
    primaryClass = "nucl-ex",
    doi = "10.1103/PhysRevLett.126.012002",
    journal = "Phys. Rev. Lett.",
    volume = "126",
    number = "1",
    pages = "012002",
    year = "2021"
}

@article{COHERENT:2024axu,
    author = "Adamski, S. and others",
    collaboration = "COHERENT",
    title = "{Evidence of Coherent Elastic Neutrino-Nucleus Scattering with COHERENT{\textquoteright}s Germanium Array}",
    doi = "10.1103/PhysRevLett.134.231801",
    journal = "Phys. Rev. Lett.",
    volume = "134",
    number = "23",
    pages = "231801",
    year = "2025"
}

@article{Ackermann:2025obx,
    author = "Ackermann, N. and others",
    title = "{Direct observation of coherent elastic antineutrino{\textendash}nucleus scattering}",
    eprint = "2501.05206",
    archivePrefix = "arXiv",
    primaryClass = "hep-ex",
    doi = "10.1038/s41586-025-09322-2",
    journal = "Nature",
    volume = "643",
    number = "8074",
    pages = "1229--1233",
    year = "2025"
}

@article{Fuss:2022fxe,
    author = "Adari, P. and others",
    editor = "Fuss, A. and Kaznacheeva, M. and Reindl, F. and Wagner, F.",
    title = "{EXCESS workshop: Descriptions of rising low-energy spectra}",
    eprint = "2202.05097",
    archivePrefix = "arXiv",
    primaryClass = "astro-ph.IM",
    reportNumber = "FERMILAB-CONF-22-208-PPD-SCD-V",
    doi = "10.21468/SciPostPhysProc.9.001",
    journal = "SciPost Phys. Proc.",
    volume = "9",
    pages = "001",
    year = "2022"
}

@article{Angloher_2026,
   title="{Optimization of TES Design for the CRESST Experiment}",
   volume={36},
   ISSN={2378-7074},
   url={http://dx.doi.org/10.1109/TASC.2026.3680071},
   DOI={10.1109/tasc.2026.3680071},
   number={6},
   journal={IEEE Transactions on Applied Superconductivity},
   publisher={Institute of Electrical and Electronics Engineers (IEEE)},
   author={Angloher, G. and others},
   year={2026},
    pages={1–7} 
}

@article{TESSERACT:2025tfw,
    author = "Bui, T. K. and others",
    collaboration = "TESSERACT",
    title = "{First Limits on Light Dark Matter Interactions in a Low Threshold Two-Channel Athermal Phonon Detector from the TESSERACT Collaboration}",
    eprint = "2503.03683",
    archivePrefix = "arXiv",
    primaryClass = "hep-ex",
    doi = "10.1103/hsrl-crvf",
    journal = "Phys. Rev. Lett.",
    volume = "135",
    number = "16",
    pages = "161002",
    year = "2025"
}

@article{DelCastello:2024ehl,
    author = "Del Castello, G.",
    title = "{LANTERN: A multichannel light calibration system for cryogenic detectors}",
    doi = "10.1016/j.nima.2024.169728",
    journal = "Nucl. Instrum. Methods Phys. Res. A",
    volume = "1068",
    pages = "169728",
    year = "2024"
}

@phdthesis{DelCastello:2025tyu,
    author = "Del Castello, G.",
    title = "{Calibration and commissioning results of the NUCLEUS experiment}",
    school = "Rome U.",
    year = "2025",
    url = "https://iris.uniroma1.it/handle/11573/1730718"
}

@article{Abdelhameed2020Tungsten,
  author  = {Abdelhameed, A. H. and others},
  title   = {Deposition of Tungsten Thin Films by Magnetron Sputtering for Large-Scale Production of Tungsten-Based Transition-Edge Sensors},
  journal = {Journal of Low Temperature Physics},
  year    = {2020},
  volume  = {199},
  number  = {1},
  pages   = {401--407},
  url     = {https://pure.mpg.de/rest/items/item_3192781/component/file_3202943/content}
}

@article{Wagner:2022xde,
    author = "Wagner, F. and Bartolot, D. and Rizvanovic, D. and Reindl, F. and Schieck, J. and Waltenberger, W.",
    title = "{Cait: Analysis Toolkit for Cryogenic Particle Detectors in Python}",
    eprint = "2207.02187",
    archivePrefix = "arXiv",
    primaryClass = "physics.ins-det",
    doi = "10.1007/s41781-022-00092-4",
    journal = "Comput. Softw. Big Sci.",
    volume = "6",
    number = "1",
    pages = "19",
    year = "2022"
}

@article{Abele:2026sqm,
    author = "Abele, H. and others",
    title = "{Characterization of the low energy excess using a NUCLEUS Al$_2$O$_3$ detector}",
    eprint = "2603.07687",
    archivePrefix = "arXiv",
    primaryClass = "physics.ins-det",
    doi = "10.1140/epjc/s10052-026-15950-3",
    journal = "Eur. Phys. J. C",
    volume = "86",
    number = "7",
    pages = "831",
    year = "2026"
}

@article{cresst2024,
    author = "Angloher, G. and others",
    collaboration = "CRESST",
    title = "{DoubleTES detectors to investigate the CRESST low energy background: results from above-ground prototypes}",
    eprint = "2404.02607",
    archivePrefix = "arXiv",
    primaryClass = "physics.ins-det",
    doi = "10.1140/epjc/s10052-024-13282-8",
    journal = "Eur. Phys. J. C",
    volume = "84",
    number = "10",
    pages = "1001",
    year = "2024",
    note = "[Erratum: Eur.Phys.J.C 84, 1227 (2024)]"
}

@misc{aurubis,
    author = "Aurubis",
    title = "{NOSV copper}",
    key = "https://www.aurubis.com/",
    url = "https://www.aurubis.com/",
    access = "04.04.2026"
}

@article{geant4_2003,
title = "{Geant4—a simulation toolkit}",
journal = {Nuclear Instruments and Methods in Physics Research Section A: Accelerators, Spectrometers, Detectors and Associated Equipment},
volume = {506},
number = {3},
pages = {250-303},
year = {2003},
issn = {0168-9002},
doi = {https://doi.org/10.1016/S0168-9002(03)01368-8},
url = {https://www.sciencedirect.com/science/article/pii/S0168900203013688},
author = {S. Agostinelli and others},
}

@article{geant4_2016,
title = "{Recent developments in Geant4}",
journal = {Nuclear Instruments and Methods in Physics Research Section A: Accelerators, Spectrometers, Detectors and Associated Equipment},
volume = {835},
pages = {186-225},
year = {2016},
issn = {0168-9002},
doi = {https://doi.org/10.1016/j.nima.2016.06.125},
url = {https://www.sciencedirect.com/science/article/pii/S0168900216306957},
author = {J. Allison and others}
}

@phdthesis{trunc_fit_schmaler2010,
	author = {Schmaler, Jens Michael},
	title = "{The CRESST Dark Matter Search - New Analysis Methods and Recent Results}",
	year = {2010},
	school = {Technische Universität München},
	url = {https://mediatum.ub.tum.de/998304}
}

@article{Bravin1999CRESST,
  author  = {Bravin, M. and others},
  title   = "{The {CRESST} Dark Matter Search}",
  journal = {Astroparticle Physics},
  volume  = {12},
  number  = {1},
  pages   = {107--114},
  year    = {1999},
  doi     = {10.1016/S0927-6505(99)00073-0},
  eprint  = {hep-ex/9904005},
  archivePrefix = {arXiv}
}

@article{Angloher2026CRESST,
  author  = {Angloher, G. and others},
  title   = "{The {CRESST} Experiment Towards the Next Generation
             of Sub-{GeV} Direct Dark Matter Detection}",
  journal = {Communications Physics},
  volume  = {9},
  number  = {1},
  pages   = {163},
  year    = {2026},
  doi     = {10.1038/s42005-025-02476-5}
}

\end{document}